%% file: TCAD-TSV-Flow.tex
\documentclass[journal,10pt]{IEEEtran}
\input{./ieee-normal}

\graphicspath{{./figs/}}

\newcommand{\revise}[1]{\textcolor{black}{#1}}

\begin{document}


\title{
Adaptive 3D-IC TSV Fault Tolerance Structure Generation
}

\author{
Song Chen,~\IEEEmembership{Member,~IEEE},
Qi Xu,~\IEEEmembership{Student Member,~IEEE},
Bei Yu,~\IEEEmembership{Member,~IEEE}
\thanks{This work was supported in part by the National Natural Science Foundation of China (NSFC) under grant No.~61674133,
61404123 and Anhui Provincial Natural Science Foundation (1508085MF134, China),
and The Research Grants Council of Hong Kong SAR (Project No.~CUHK24209017).}
\thanks{S.~Chen and Q.~Xu are with Department of Electronic Science and Technology, University of Science and Technology of China, China (e-mail: \href{mailto:songch@ustc.edu.cn}{\texttt{songch@ustc.edu.cn}}, \href{mailto:xuqi@mail.ustc.edu.cn}{\texttt{xuqi@mail.ustc.edu.cn}}).}
\thanks{B.~Yu is with the Department of Computer Science and Engineering, The Chinese University of Hong Kong, NT, Hong Kong (e-mail: \href{mailto:byu@cse.cuhk.edu.hk}{\texttt{byu@cse.cuhk.edu.hk}}).}
}

\markboth{IEEE TRANSACTIONS ON COMPUTER-AIDED DESIGN OF INTEGRATED CIRCUITS AND SYSTEMS, VOL. , NO.}{Song Chen et al.}

\maketitle
\thispagestyle{empty}

\input{doc/abstract}
\input{doc/intro}

\input{doc/problem}

\input{doc/structure}

\input{doc/ilp}
\input{doc/heuristic}

\input{doc/flow}
\input{doc/result}

\input{doc/conclu}


\input{TCAD-TSV-Flow.bbl}


\end{document}

%% file: ieee-normal.tex
\usepackage{cancel}
\usepackage{algpseudocode}                                 
\usepackage{algorithm}
\usepackage{graphicx}                                      
\usepackage{amsmath}
\usepackage{amssymb}
\usepackage{amsfonts}
\usepackage{booktabs}
\usepackage{multirow}
\usepackage[subrefformat=parens,farskip=0pt,justification=centering]{subfig}
\usepackage{color, colortbl}
\usepackage{cite}                                          
\usepackage{comment}                                       
\usepackage{soul}                                          
\soulregister\cite7
\soulregister\ref7
\soulregister\pageref7
\usepackage{amsthm}
\usepackage{etoolbox}                                      
\usepackage{url}
\usepackage{nth}                                           
\usepackage{bm}                                            
\usepackage{courier}
\usepackage{balance}
\usepackage{threeparttable}
\usepackage[bookmarks=false]{hyperref}                     %
\hypersetup{
    colorlinks = true,
    citecolor  = blue,
    linkcolor  = blue,
    urlcolor   = blue,
}
\usepackage{filecontents}                                  
\usepackage{pgfplots}
\usepackage{pgfplotstable}
\pgfplotsset{compat=newest}
\usepackage{caption}
\pgfplotsset{compat=newest}
\newtheorem{myproblem}{\textbf{Problem}}

\newtheorem{mytheorem}{\textbf{Theorem}}
\newtheorem{mylemma}{\textbf{Lemma}}

\algrenewcommand\textproc{\texttt}
\newcommand{\tabincell}[2]{
    \begin{tabular}{@{}#1@{}}#2\end{tabular}
}

\makeatletter
\let\OldStatex\Statex
\renewcommand{\Statex}[1][3]{%
  \setlength\@tempdima{\algorithmicindent}%
  \OldStatex\hskip\dimexpr#1\@tempdima\relax
}
\makeatother

\RequirePackage[normalem]{ulem} 
\RequirePackage{color}\definecolor{RED}{rgb}{1,0,0}\definecolor{BLUE}{rgb}{0,0,1} 

%% file: doc/abstract.tex
\begin{abstract}
In three dimensional integrated circuits (3D-ICs), through silicon via (TSV) is a critical technique in providing vertical connections.
However, the yield and reliability is one of the key obstacles to adopt the TSV based 3D-ICs technology in industry.
Various fault-tolerance structures using spare TSVs to repair faulty functional TSVs have been proposed in literature for yield and reliability enhancement,
but a \revise{valid} structure cannot always be found due to \revise{the lack of effective generation methods for fault-tolerance structures}.
In this paper, we focus on the problem of adaptive fault-tolerance structure generation.
Given the relations between functional TSVs and spare TSVs, we first calculate the maximum number of tolerant faults in each TSV group.
Then we propose an integer linear programming (ILP) based model to construct adaptive fault-tolerance structure with minimal multiplexer delay overhead and hardware cost.
We further develop a speed-up technique through efficient min-cost-max-flow (MCMF) model.
All the proposed methodologies are embedded in a top-down TSV planning framework to form functional TSV groups and generate adaptive fault-tolerance structures.
Experimental results show that, compared with state-of-the-art, the number of spare TSVs used for fault tolerance can be effectively reduced. 

\end{abstract}

%% file: doc/intro.tex
\section{Introduction}
\label{sec:intro}

\IEEEPARstart{A}{s} device feature sizes continue to rapidly decrease, the interconnect delay is becoming a bottleneck limiting IC performance.
Three dimensional integrated circuits (3D-ICs) technology involves vertically stacking multiple dies connected by through silicon vias (TSVs),
providing a promising way to alleviate the interconnect problem and achieve a significant reduction in chip area, wire-length and interconnect power \cite{souri2000multiple}.
Study indicates that the average wire-length of a 3D-IC varies according to the square root of the number of layers \cite{joyner2001global}.
Moreover, 3D-ICs also offer the potential for heterogeneous integration, which is essential for More than Moore (MtM) technology \cite{ITRS}.
3D integration has already seen commercial applications in the form of 3D memory but there are still significant open problems in both research and implementation \cite{tcad2017Lu}. In this work, we will focus on the TSV reliability problem.

TSVs may be affected by various reliability issues such as undercut, misalignment, or random open defects~\cite{loi2008low}.
Because there exist a large number of TSVs in a chip, these issues in turn lead to low chip yield.
For example,~\cite{loi2008low,chen2015novel} reported a 60\% chip yield for a chip with 20000 TSVs and only 20\% yield for 55000 TSVs in IMEC process technology.
Since yield and reliability is a primary concern in 3D ICs design, a robust fault-tolerance structure is imperative.
In general, there are two types of yield losses in 3D-ICs: the yield loss due to defects in stacked dies and the yield loss due to defects occurred during assembling process \cite{xu2012yield}.
For the former case, it is critical to conduct pre-bond testing to avoid the stacking of defective dies \cite{lee2009test}.
A number of die/wafer matching and inter-die repair strategies have also been proposed to increase the stack yield \cite{ferri2007strategies,chou2010yield,jiang2010yield,jiang2012effective}.
For the latter case, adding {spare TSVs} (referred to as \textbf{s-TSVs}) to repair fault {functional TSVs} (referred to as \textbf{f-TSVs}) is an effective method for enhancing yield.

One key problem in TSV fault-tolerance design is the fault-tolerance structure generation, where a number of functional TSVs and one or several spare TSVs are grouped together to provide redundancy.
Chen \textit{et al}.~\cite{chen2015novel} proposed a minimum spanning tree based method to group f-TSVs and form one-fault-tolerance structures.
However, the method is difficult to be applied to multiple-fault-tolerance structure generation.
Wang \textit{et al}.~\cite{wang2015defect} presented a regular TSV replacing chain structure that can repair faulty TSVs based on a realistic clustered defect model.
Xu \textit{et al}.~\cite{xu2017clustered} further considered the physical information of the TSV groups,
and developed an ILP formulation for fault-tolerance structure generation.
They model replaceable relations between f-TSVs, so the maximum input-port number of individual multiplexers can be effectively reduced.
However, all previous works \cite{wang2015defect,xu2017clustered} are under an assumption that a predetermined number of s-TSVs are assigned to each TSV group.
To ensure that $K$ common s-TSVs can be allocated to each f-TSV group, in each group f-TSV number is usually quite small, which introduces a large number of TSV groups.
Since the total number of s-TSVs is proportional to the TSV group number, it may cause overuse of s-TSVs.

\begin{figure*}[tbp!]
    \centering
    \subfloat[]{\includegraphics[height=4.2cm]{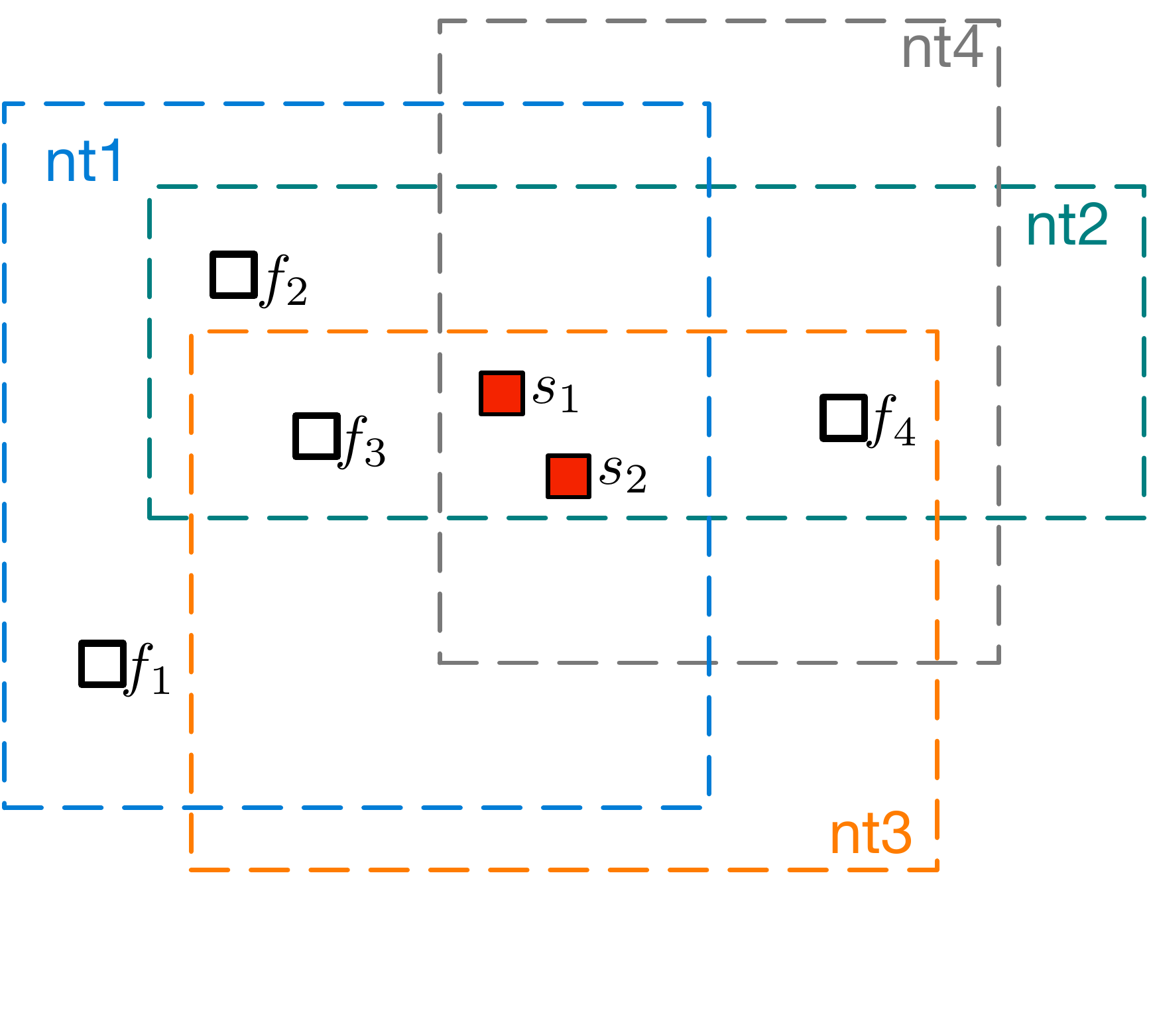}  \label{fig:struct:a}}
    \subfloat[]{\includegraphics[height=5.0cm]{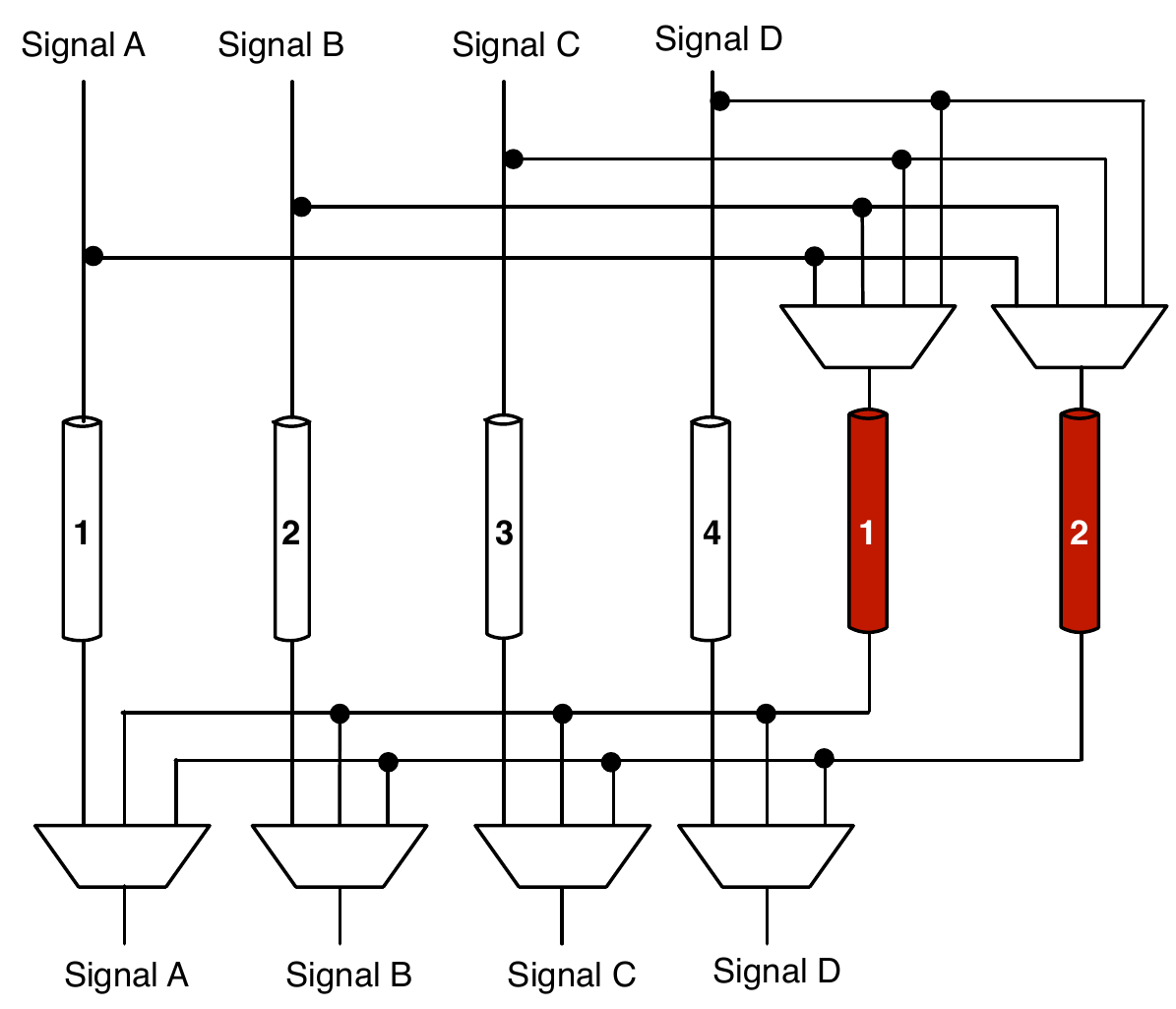}  \label{fig:struct:b}}
    \subfloat[]{\includegraphics[height=5.0cm]{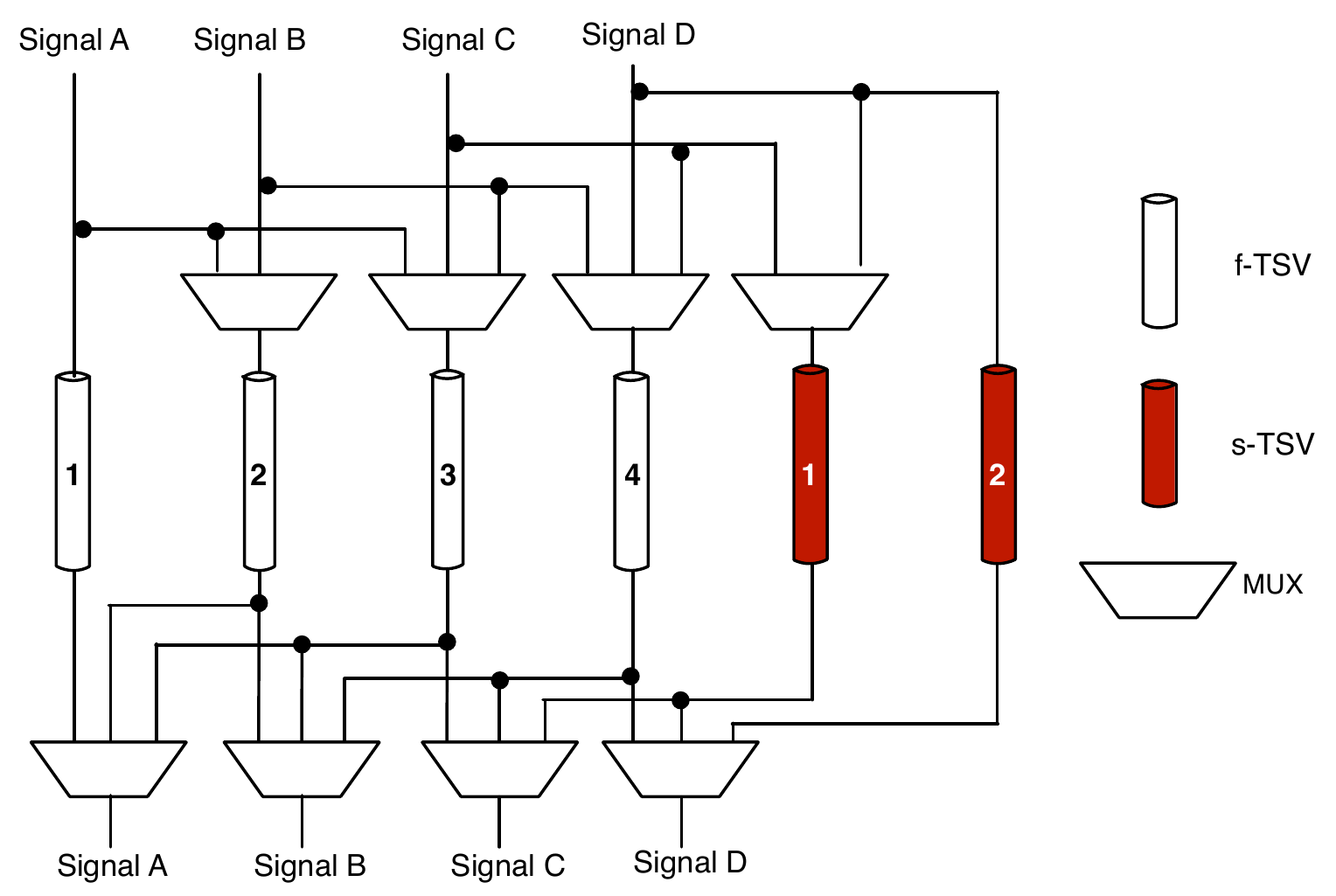}  \label{fig:struct:c}}
    \caption{
        (a) An example of TSV group with four f-TSVs and two s-TSVs;
        (b) A fault-tolerance structure with large multiplexer delay overhead;
        (c) A regular chain structure.
    }
    \label{fig:struct}
\end{figure*}

To overcome the above issue, in this paper we propose an adaptive fault-tolerance structure,
in which the number of tolerant faults is adaptively determined by the distribution of the f-TSVs and their candidate s-TSVs.
A set of s-TSVs will be selected from a large amount of candidates.
Our adaptive fault-tolerance structure generation method can achieve minimal multiplexer delay overhead, as well as minimal number of required s-TSVs.
Key technical contributions of this work are listed as follows.
\begin{itemize}
  \item We are able to determine the maximum number of tolerant faults, denoted as $K$, in polynomial time.
  \item We present an integer linear programming formulation in generating the adaptive $K$-fault tolerance structures. 
  \item We further propose an efficient min-cost-max-flow (MCMF) based heuristic method to speed-up the $K$-fault tolerance structure generation.
  \item All the proposed methodologies are embedded in a top-down TSV planning framework to form f-TSV groups and generate fault-tolerance structures.

\end{itemize}

Experimental results show that, compared with state-of-the-art, the proposed framework can reduce the number of used s-TSVs and maximum port number of multiplexers. 


The remainder of this paper is organized as follows.
Section \ref{sec:background} presents the motivation and gives the problem formulation. 
The method for determining the maximum number of tolerant faults is presented in Section \ref{sec:ta}.
Section \ref{sec:ilp} and Section \ref{sec:hm} present the proposed ILP formulation and heuristic method. 
Section \ref{sec:mgy} describes the proposed fault tolerance TSV planning methodology.
Section \ref{sec:result} provides experimental results, followed by conclusion in Section \ref{sec:conclu}.

%% file: doc/problem.tex
\section{Preliminaries}
\label{sec:background}

\subsection{Chip Yield and TSV Yield}
\label{subsec:cty}
Consider a 3D IC containing $l$ layers, and the yield of $i^{th}$ layer die is $Y_{die_i}$. The yield for wafer-to-wafer (W2W) stacking $Y_{stack}$ can be roughly modeled as~\cite{xu2012yield}:
\begin{equation}
   \label{eq:yield}
   Y_{stack} = \prod_{i=1}^{l}(Y_{die_i})
\end{equation}
Therefore,
the defects exist in each die will certainly affect the overall chip yield after stacking.

Besides, during bonding, any foreign particle caught between the wafers can lead to peeling, as well as delamination, which dramatically reduces bonding quality and yield~\cite{chen2010cost}. $Y_{Bonding}$ captures the yield loss of the chip due to faults in the bonding processes.

According to the cumulative yield property, the yield of a 3D chip $Y_{3D-chip}$ can be formulated as follows \cite{xu2012yield}:

\begin{align}
Y_{3D-chip}=Y_{stack}\cdot\prod_{i=1}^{l-1}(Y_{Bonding(i)}\cdot Y_{TSV(i)}),
\end{align}
where $Y_{Bonding(i)}$ is the yield of the $i^{th}$ bonding step, and $Y_{TSV(i)}$ is the TSV yield in the $i^{th}$ layer.
In our work, we focus on the yield enhancement of 3D chip in terms of TSV yield $Y_{TSV}$ \cite{wang2015defect}.
The total TSV yield $Y_{TSV}$ is calculated by multiplying all f-TSV group yield $Y_{gj}$ as follows.
\begin{align}
Y_{TSV}=\prod_{j=1}^{N}{Y_{gj}},
\end{align}
where $N$ is the number of f-TSV groups.
In this paper we adopt the algorithm described in \cite{wang2015defect} for the calculation of group yield $Y_{gj}$.

\begin{figure*}[tbp!]
    \centering
    \subfloat[]{\includegraphics[height=4.8cm]{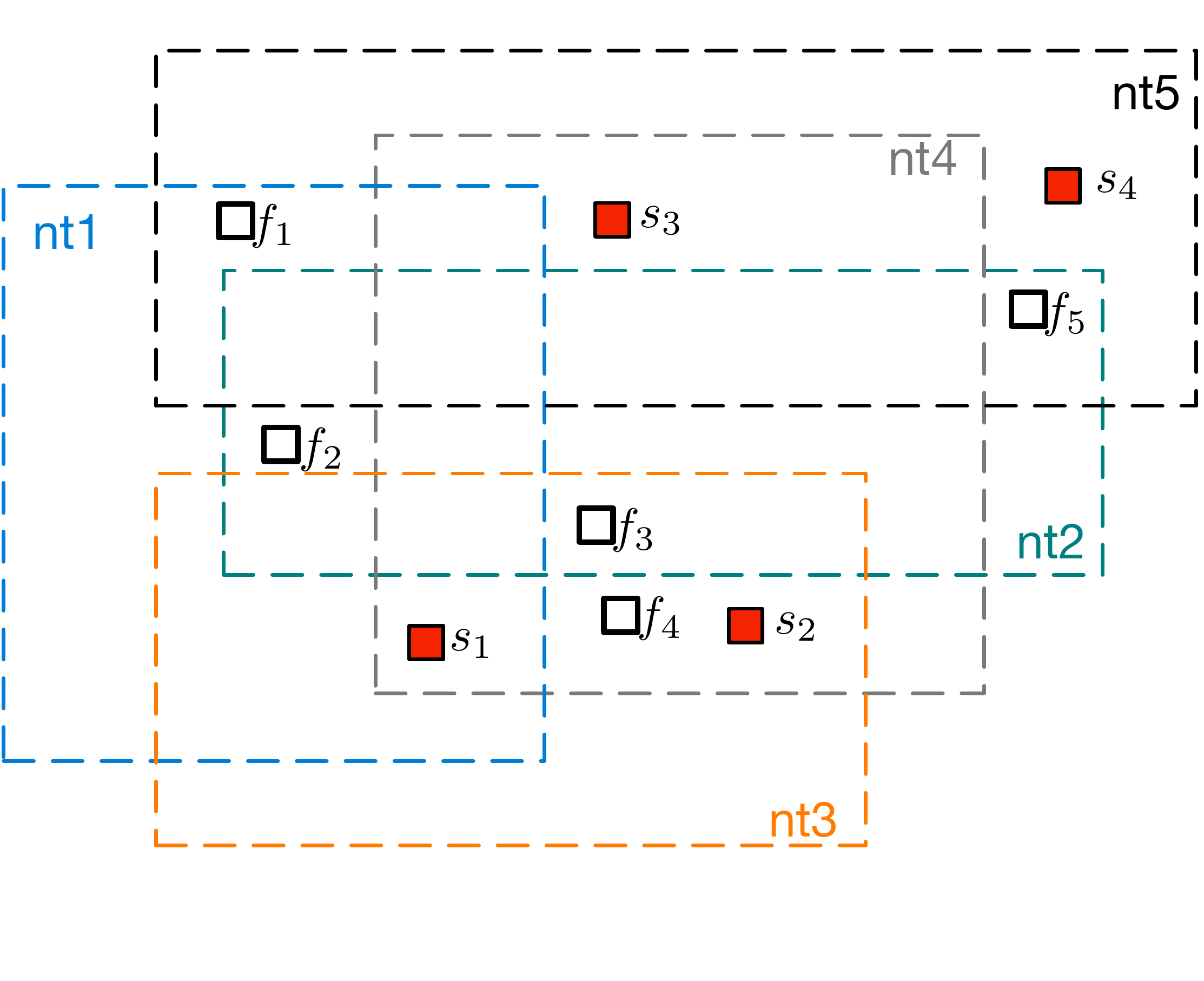}  \label{fig:special:a}}
    \hspace{.1in}
    \subfloat[]{\includegraphics[height=5.4cm]{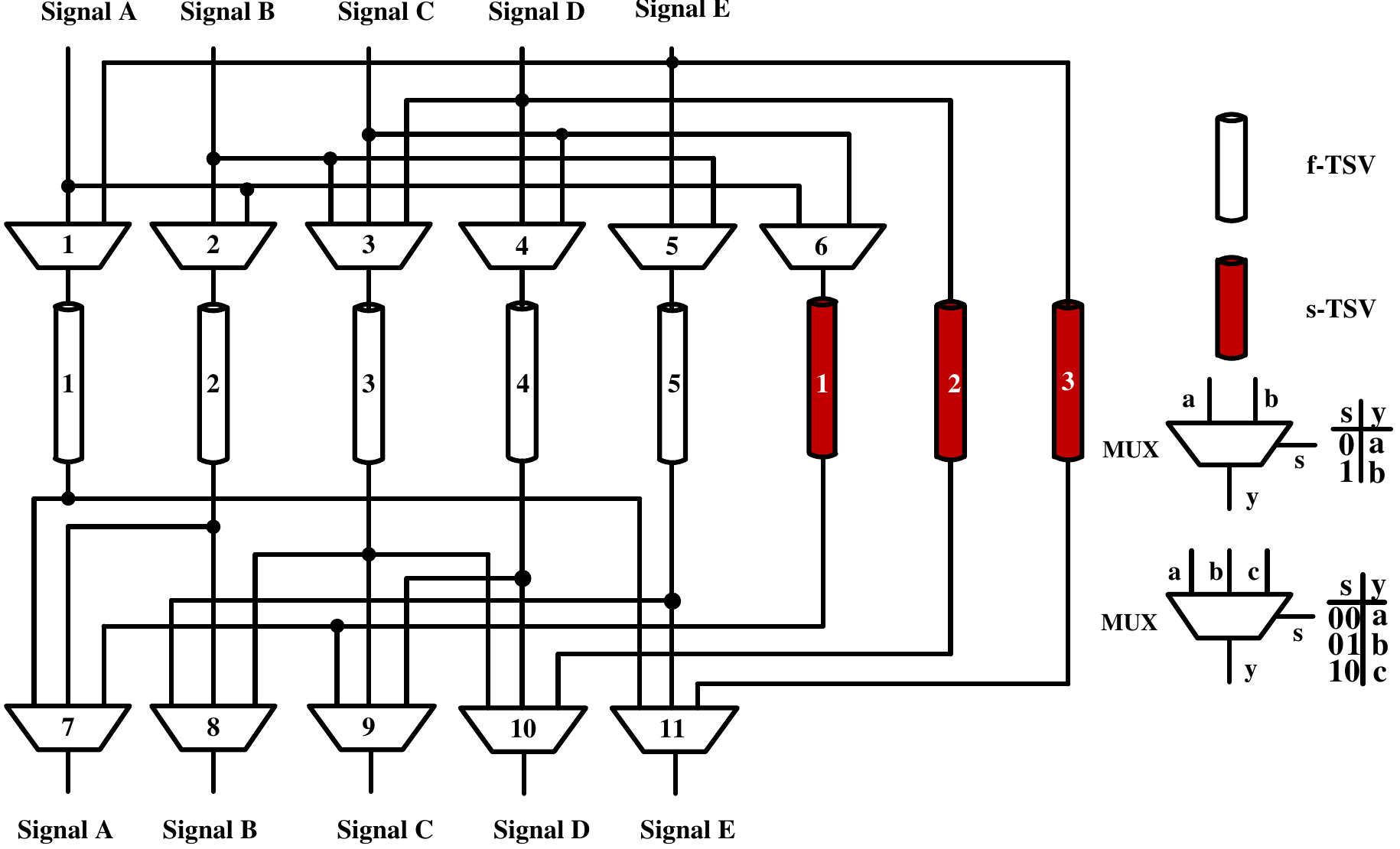}  \label{fig:special:b}}
    \hspace{.1in}
    \caption{(a) An example of TSV group with five f-TSVs and four s-TSVs, which cannot be handled by previous works; (b) The adaptive fault tolerance structure generated by our proposed methodology.}
    \label{fig:special}
\end{figure*}

\subsection{TSV Fault-Tolerance Structure}
\label{sec:tfts}

By inserting the multiplexers (including control circuits) and carefully designing the reconfigurable TSV replacing paths, we can construct TSV fault-tolerance structures, where the s-TSVs can be used to transfer signals in the presence of faulty f-TSVs \cite{loi2008low}.

Given an f-TSV planning result, we know the number and positions of all f-TSVs.
Then we perform a top-down iterative f-TSV partitioning to form f-TSVs groups and allocate s-TSVs in the whitespace for each group.
The number and positions of used s-TSVs for each f-TSV group are determined simultaneously in the f-TSV partitioning stage.
Fig.~\subref*{fig:struct:a} shows an example of a TSV group with four f-TSVs ($f_1 \cdots f_4$) and two s-TSVs ($s_1$ and $s_2$).
Here $f_1 \cdots f_4$ belong to nets $nt_1 \cdots nt_4$, respectively.
The dashed large rectangles represent the bounding boxes of different nets.
Without loss of generality, we denote the bounding box of an f-TSV $f_i$ as the bounding box of the net $f_i$ belonging to.
We say that an f-TSV $f_i$ can be replaced by another TSV $v$, if and only if $v$ is located inside or nearby the bounding box of $f_i$. 
Note that here the TSV $v$ can be either f-TSV or s-TSV.
For example, $f_1$ is replaceable by $f_2$, $f_3$, $s_1$, $s_2$, since these four TSVs are covered by the bounding box of $f_1$.

Given a TSV group with some f-TSVs and \textcolor{blue}{$K$} s-TSVs, a $K$-fault tolerance structure includes $K$ independent directed TSV-replacing paths from each f-TSV to s-TSVs.
In this structure we can repair at most $K$ faulty f-TSVs through multiplexer rerouting.
For instance, for the TSV group shown in Fig.~\subref*{fig:struct:a}, a 2-fault tolerance structure with two s-TSVs can be generated as in Fig.~\subref*{fig:struct:b},
where each f-TSV is directly connected to all s-TSVs.
Although the design scheme is very simple, this structure suffers from large delay overhead due to large multiplexer input size.
Some recent works \cite{wang2015defect,xu2017clustered} proposed regular $K$-fault tolerance structure, as shown in Fig.~\subref*{fig:struct:c}.
Here each f-TSV is regularly connected to two right side neighbouring TSVs and the rightmost f-TSVs are connected to s-TSVs.
Instead of 4-port multiplexers occupied in Fig.~\subref*{fig:struct:b}, here only 3-port multiplexers and 2-port multiplexers are needed.
For each f-TSV, the independent TSV-replacing paths are listed as follows.
\begin{description}
  \item[$f_1$:] \{$f_1 \rightarrow f_3 \rightarrow s_1$\}, \{$f_1 \rightarrow f_2 \rightarrow f_4 \rightarrow s_2$\}.
  \item[$f_2$:] \{$f_2 \rightarrow f_3 \rightarrow s_1$\}, \{$f_2 \rightarrow f_4 \rightarrow s_2$\}.
  \item[$f_3$:] \{$f_3 \rightarrow s_1$\}, \{$f_3 \rightarrow f_4 \rightarrow s_2$\}.
  \item[$f_4$:] \{$f_4 \rightarrow s_1$\}, \{$f_4 \rightarrow s_2$\}.
\end{description}

\begin{figure*}[tb!]
    \centering
    \subfloat[]{\includegraphics[height=3.0cm]{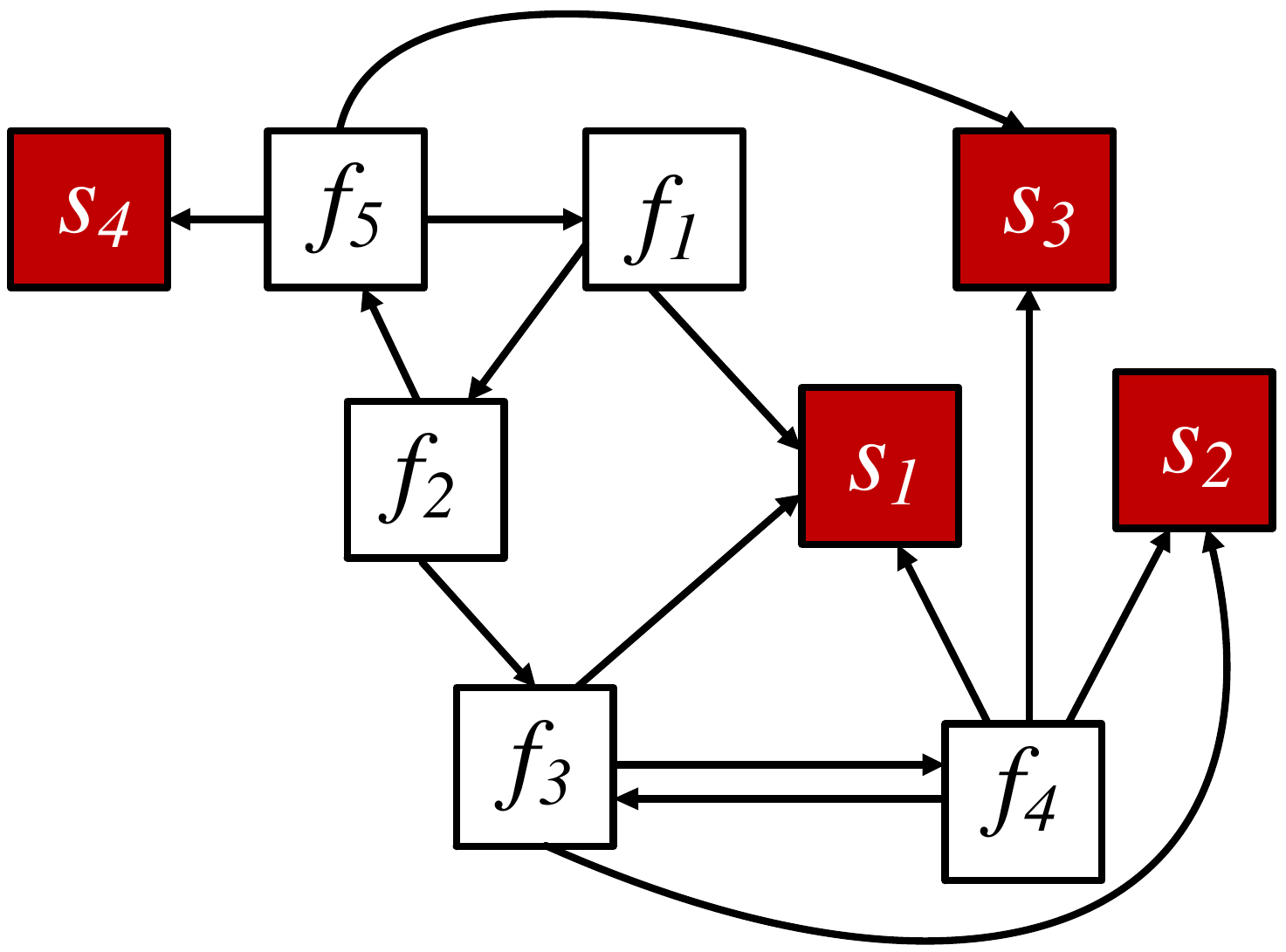}       \label{fig:origgraph1}}
    \hspace{.1in}
    \subfloat[]{\includegraphics[height=3.0cm]{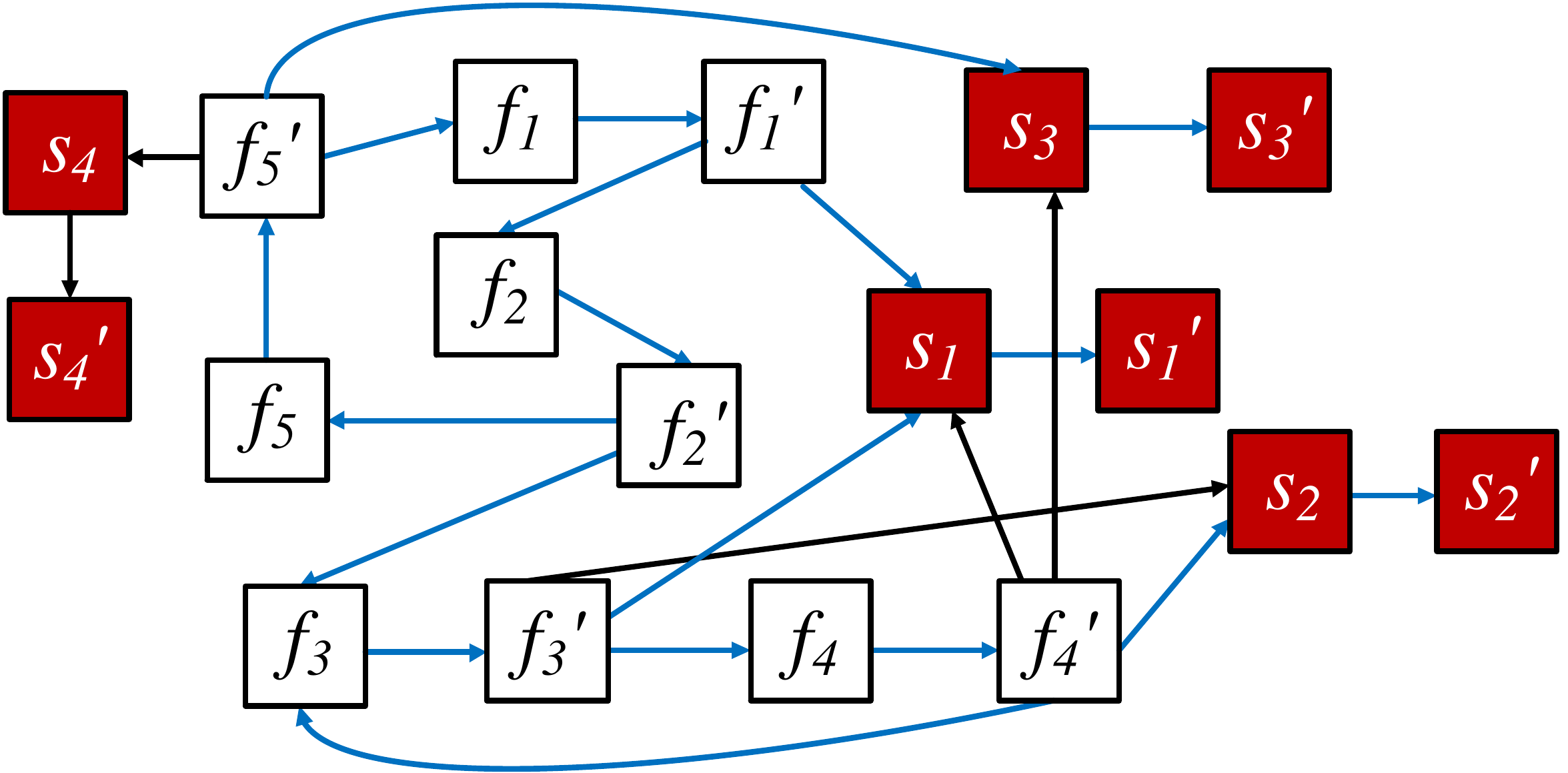}       \label{fig:splitgraph}}
    \hspace{.1in}
    \subfloat[]{\includegraphics[height=3.0cm]{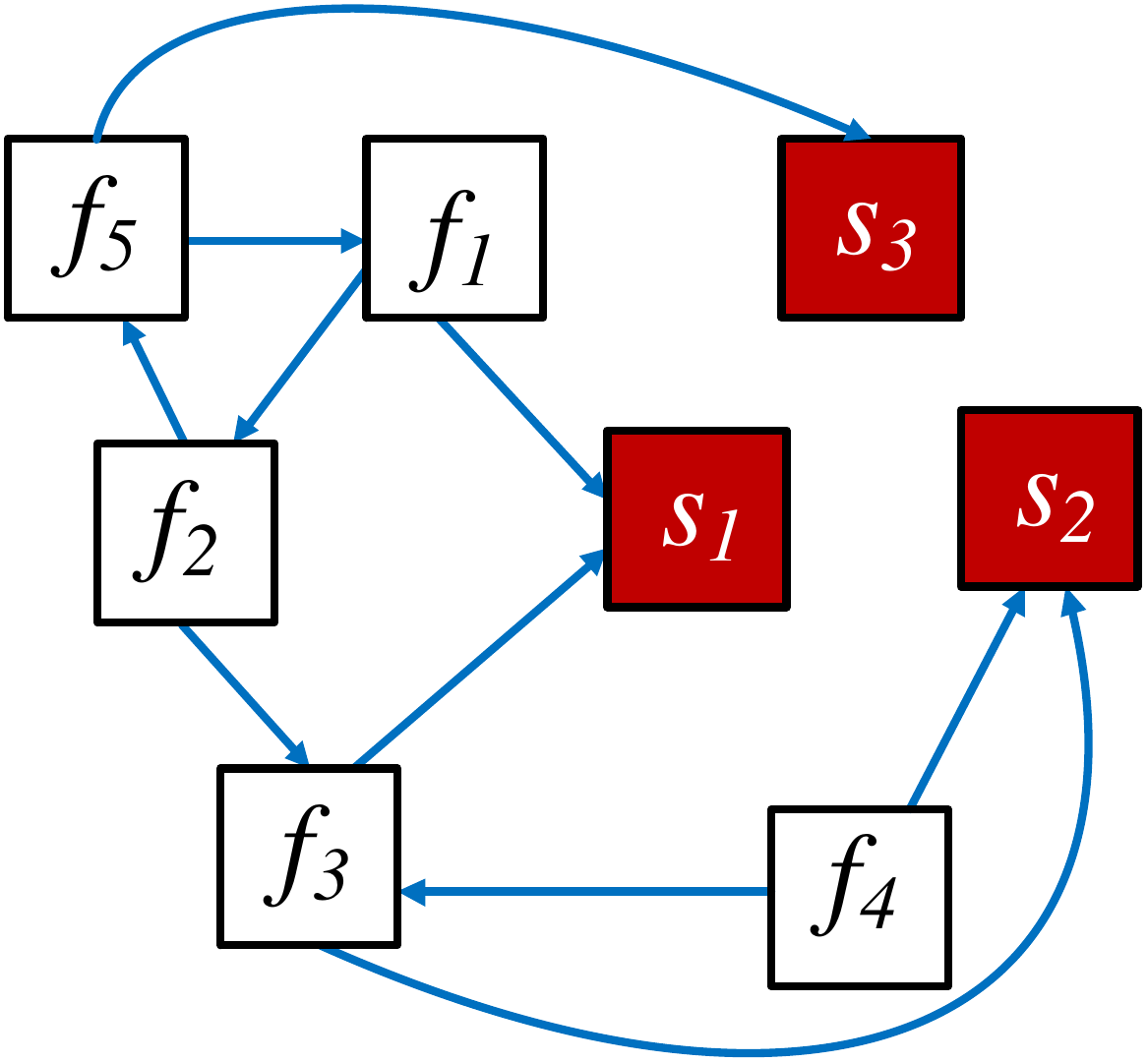}       \label{fig:origgraph2}}
    \caption{(a) The corresponding directed graph $G$ of layout in Fig. 2(a); (b)The corresponding splitting graph $G'$; (c) 2-fault tolerance structure on graph $G$.}
    \label{fig:origraph}
\end{figure*}

\revise{To ensure the existence of fault-tolerance structures in TSV groups, the previous works (e.g.~\cite{wang2015defect,xu2017clustered}) form TSV groups under two constraints: (1) $K$ fault-tolerance structures use exactly $K$ s-TSVs and (2) an f-TSV in a group can be replaced by any s-TSV within the group.
Fig.~\subref*{fig:struct:a} shows an example of TSV group having two-fault tolerance structures, where all the f-TSVs, $f_1$, $f_2$, $f_3$, and $f_4$, can be replaced by both $s_1$ and $s_2$ considering the net bounding boxes.
Unfortunately, general cases may violate these constraints. Fig.~\subref*{fig:special:a} shows a generalized example, where five f-TSVs ($f_1 \cdots f_5$) and four s-TSVs ($s_1 \cdots s_4$) are involved. The replaceable relations between TSVs are shown in Fig.~\subref*{fig:origgraph1}. In this TSV group, the constraint (1) is violated since we cannot find two-fault tolerance structures if only two s-TSVs are used. The constraint (2) is also violated even if the group is partitioned into smaller groups since $f_2$ have no replaceable s-TSVs. Consequently, the method in \cite{wang2015defect} cannot generate cost-effective fault-tolerance structures for this TSV group, because $f_2$ has no candidate s-TSVs. The ILP-based method in \cite{xu2017clustered} cannot generate fault-tolerance structures for this TSV group since the number of tolerant faults is unknown.
However, the f-TSV group definitely includes a two-fault tolerance structure as shown in Fig.~\subref*{fig:origgraph2}, where three out of four s-TSVs are used in the fault-tolerance structure.}
The possible TSV replacing paths are as follows.
\begin{description}
    \item[$f_1$:] $\{f_1 \rightarrow s_1\}, \{f_1 \rightarrow f_2 \rightarrow f_3 \rightarrow f_4 \rightarrow s_2\}$.
    \item[$f_2$:] $\{f_2 \rightarrow f_5 \rightarrow f_1 \rightarrow s_1\}, \{f_2 \rightarrow f_3 \rightarrow f_4 \rightarrow s_2\}$.
    \item[$f_3$:] $\{f_3 \rightarrow s_1\}, \{f_3 \rightarrow f_4 \rightarrow s_2\}$.
    \item[$f_4$:] $\{f_4 \rightarrow f_3 \rightarrow s_1\}, \{f_4 \rightarrow s_2\}$.
    \item[$f_5$:] $\{f_5 \rightarrow f_1 \rightarrow s_1\}, \{f_5 \rightarrow s_3\}$.
\end{description}

In reality, there is no essential difference between the f-TSVs and s-TSVs. Therefore, the existing TSV testing technique can be directly adopted to test the f-TSVs and s-TSVs~\cite{brandon2014design}.
And the control signal of multiplexers can be set to determine the direction of signal transfer.
As shown in Fig.~\subref*{fig:special:b}, the control signal of 2-to-1 and 3-to-1 multiplexer are 1-bit and 2-bit, respectively.
When all TSVs are fault-free or existing faulty s-TSVs, the control signals of each multiplexer are set to transfer signal through their corresponding f-TSVs.
But once an f-TSV is faulty, the reconfigurable routing paths can be determined by the corresponding control signal of multiplexers.
For instance, when f-TSV 1 is faulty, the control signals of multiplexer 6 and 7 are set to 0 and 10, causing s-TSV 1 to reroute the signal $A$.

\subsection{Hardware Cost and Multiplexer Delay Overhead}
\label{sec:mdo}
The hardware cost incurred by the fault-tolerance structure can be divided into several parts, including the area overhead due to inserted s-TSVs, related control logic (i.e., MUXes), and re-routing interconnect~\cite{wang2015defect}. And the cost is dominated by the first two parts~\cite{jiang2012effective}.
Jiang \textit{et al}.~\cite{jiang2013effective} point out that the area of control logic is negligible compared with the TSV size and the TSV manufacturing cost is much larger than logic gates.
Therefore, in order to reduce the hardware cost, we should reduce the number of s-TSVs used in the fault-tolerance structures.

The delay of a multiplexer is increased along with the number of ports. Therefore, a large multiplexer will introduce large delay overhead.
Moreover, the proposed TSV fault tolerance planning is performed in floorplanning stage and we have no exact timing information.
If we minimize the multiplexer delay overhead in this stage, we could alleviate the timing closure issue in next placement and routing stage.
Therefore, in our work, we consider the multiplexer delay overhead as one of the optimization objectives.


\subsection{Problem Formulation}
\label{sec:problem}

From the example in Fig.~\ref{fig:special}, we can see that we confront new design challenging if not all s-TSVs can be occupied in constructing $K$-fault tolerance structure.
Given a TSV group with $m$ f-TSVs and $n$ s-TSVs, we first construct a directed graph $G (V, E)$ consisting of all TSV replaceable relations.
Here vertex set $V = V_1 \cup V_2$, where $V_1$ = $\{f_i|i=1,\cdots,m\}$ is the f-TSVs set and $V_2$ = $\{s_i|i=1,\cdots,n\}$ is the s-TSVs set.
Besides, the edge set $E = \{(u, v)|{u \in V_1} \wedge v\in V \wedge u \textrm{ can be replaced by } v\}$.
Given the TSV group in Fig.~\subref*{fig:special:a}, the corresponding replaceable relation graph is shown in Fig.~\subref*{fig:origgraph1}.

We define the problem of \textbf{TSV fault-tolerance structure generation} as follows.

\begin{myproblem}
Given a TSV group with $m$ f-TSVs and $n$ s-TSVs, and the directed graph $G (V, E)$, we search for the maximum number of tolerant faults $K$.
Then we generate a $K$-fault tolerance structure, which includes $K$ independent TSV replacing paths (vertex-disjoint) for each f-TSVs, to minimize both the multiplexer delay overhead and the number of used s-TSVs.
\end{myproblem}
\revise{
Notice that the yield of the TSV group is evaluated based on the allocated s-TSVs and the f-TSVs. With the yields of the TSV groups, the total TSV yield can be calculated as discussed in Section \ref{subsec:cty}. If the target TSV yield is not satisfied, a TSV group will be selected and partitioned into two smaller new TSV groups, where the above \textbf{TSV fault-tolerance structure generation} problem will be solved again. New TSV groups will be iteratively generated until the target chip yield is satisfied.}





%% file: doc/structure.tex
\section{Max Flow based Methodology}
\label{sec:ta}

Given a TSV group with replaceable relation graph $G$, we say the TSV group has a $K$-fault tolerance structure if each f-TSV $f \in V_1$ has $K$ paths to s-TSV vertices in $G$.
Besides, for each f-TSV $f$, the paths are \textbf{vertex-disjoint} except the $f$ itself.
In this section, we develop a polynomial time algorithm to determine the $K$ value in a TSV group.
Our methodology is based on the Menger's theorem as follows.

\begin{mylemma}[Menger's theorem \cite{schrijver}]
\label{lem:lem1}
Let $G$ be a directed graph, and let $S$ and $T$ be distinct vertices in $G$. Then the maximum number of vertex-disjoint $S$-$T$ paths is equal to the minimum size of an $S$-$T$ disconnecting vertex set.
\end{mylemma}

Here the $S$-$T$ disconnecting vertex set represents a vertex set whose removal will cause no paths from any vertex in $S$ to any vertex in $T$.
According to Lemma \ref{lem:lem1}, for each f-TSV $f$, the number of vertex-disjoint paths $Nd(f)$ equals to the minimum size of the \{$f$\}-$V_2$ disconnecting vertex set in $G$.
For example, in Fig.~\subref*{fig:origgraph1}, \{$f_2,s_1$\} is a minimum \{$f_1$\}-$V_2$ disconnecting vertex set.
Therefore, the number of vertex-disjoint paths, $Nd(f_1)$, equals to 2.
Based on above lemma, we reach the following theorem:

\begin{mytheorem}
Given the replaceable relation graph, the maximum number of tolerant faults, $K$, can be determined in polynomial time, as follows:
\begin{equation}
   \label{eq:tfts}
   \textrm{K} = \mathop{\min}\limits_{f\in{V_1}}\{Nd(f)\}.
\end{equation}
\label{them:them1}
\end{mytheorem}

Since vertex-disjoint problem is not easy to model, we perform vertex splitting on $G (V, E)$ so that it can be transformed to an \textbf{edge-disjoint} problem,
which can be appropriately modelled in a maximum flow problem.
Each vertex $u \in V$ is split into two vertices $u$ and $u'$, respectively, corresponding to the vertex's input and output, and an extra edge $(u,u')$ with zero cost is also added.
A new directed graph $G'(V',E')$ is constructed as follows.
\begin{itemize}
    \item
    The vertex set $V' = V \cup V_1' \cup V_2'$, where $V_1'$ is the split vertex set of $V_1$ and $V_2'$ is the split vertex set of $V_2$.
    \item
    The edge set $E' = E_1' \cup E_2'$, where $E_1' = \{(u, u')| u \in {V} \wedge{u'}$ is the corresponding split vertex of $u$\} and $E_2' = \{(u', v)|(u,v)\in E(G) \wedge{u'}$ is the corresponding split vertex of $u$\}.
    If there is a directed edge from $u$ to $v$ in $E(G)$, a corresponding directed edge from $u'$ to $v$ is added in $E'(G')$.
\end{itemize}

Based on the splitting graph, the maximum number of tolerant faults $K$ can be determined in polynomial time by solving a max-flow problem \cite{schrijver} for each f-TSV.
For instance, given the replaceable relation graph $G (V, E)$ in Fig.~\subref*{fig:origgraph1},
Fig.~\subref*{fig:splitgraph} illustrates the splitting graph $G' (V', E')$.
The number of edge-disjoint paths for each f-TSV are as follows, $Nd(f_1)=2$, $Nd(f_2)=2$, $Nd(f_3)=3$, $Nd(f_4)=3$ and $Nd(f_5)=3$.
Since $f_1$ and $f_2$ have only two edge-disjoint paths, the maximum number of tolerant faults, $K$, equals to 2.

The fault-tolerance structure can be generated by finding $m \times K$ paths, which begin with each split f-TSV in $V_1'$ and end with split s-TSV in $V_2'$.
In addition, all the paths sharing one same source vertex should be edge-disjoint.
In the next two sections, we will propose an ILP based algorithm and a min-cost max-flow based heuristic method to generate the $K$-fault tolerance structure in minimizing both the used s-TSV number and the multiplexer delay overhead.


%% file: doc/ilp.tex
\section{Integer Linear Programming Formulation}
\label{sec:ilp}

In this section, we discuss how the $K$ edge-disjoint path search problem can be formulated as an integer programming.
For convenience, some notations used in this section are listed in TABLE~\ref{tab:tab5}.

\begin{table}[!tb]
\renewcommand{\arraystretch}{1.44}
\centering
\caption{Notations used in ILP.}
\label{tab:tab5}
    \begin{tabular}{|c|p{6.8cm}|}
    \hline
%
    $V$, $V'$        & set of f-TSVs and s-TSVs, set of split f-TSVs and split s-TSVs           \\\hline
    $V_1$, $V_1'$    & set of f-TSVs, set of split f-TSVs                 \\\hline
    $V_2$, $V_2'$    & set of s-TSVs, set of split s-TSVs                 \\\hline
    $f_i$, $f_i'$    & f-TSV in $V_1$, split f-TSV in $V_1'$              \\\hline
    $s_j$, $s_j'$    & s-TSV in $V_2$, split s-TSV in $V_2'$              \\\hline
    $E'$             & set of all edges in graph $G'$                     \\\hline
    $E_1'$           & set of all splitting edges in graph $G'$ ($f_i\rightarrow f_i'$ and $s_j\rightarrow s_j'$)         \\\hline
    $E_2'$           & set of all replaceable edges in graph $G'$         \\\hline
    $(w,w')$         & edge in $E_1'$ and $w$ in $V_2$                    \\\hline
    $s$, $t$         & split f-TSV in $V_1'$, split s-TSV in $V_2'$       \\\hline
    $v^{(s,t)}$      & binary variable; if a unit flow (path) exists from $s$ to $t$ then $v^{(s,t)} = 1$, otherwise $v^{(s,t)} = 0$\\\hline
    $(v,u)$          & edge in $E'$\\\hline
    $x_{vu}^{(s,t)}$ & binary variable; if a unit flow (path) from $s$ to $t$ goes through edge $(v,u)$, then $x_{vu}^{(s,t)} = 1$, otherwise $x_{vu}^{(s,t)} = 0$ \\\hline
    $d_{vu}$ & binary variable on edge $(v,u)$;  if a unit flow (path) goes through edge $(v,u)$, then $d_{vu} = 1$, otherwise $d_{vu} = 0$          \\\hline
\end{tabular}
\end{table}

First, an integer programming formulation in~\cite{xu2017clustered} is given to generate the fault-tolerance structures with minimization of the multiplexer delay overhead.

To model the delay of each multiplexer, it is of importance calculating indegree of each vertex $u \in V$.
As shown in Fig.~\subref*{fig:splitgraph}, the edge $(f_2',f_3)$ is on the path from $f_1'$ to $s_2'$, as well as the path from $f_2'$ to $s_2'$.
Although the same edge is traversed by two paths, it only increases the indegree of $f_3$ by one.
Meanwhile, there may be several edges directed into same TSV vertex on the paths.
For instance, due to edges $(f_2', f_3)$ and $(f_4', f_3)$, the indegree of $f_3$ should be increased by two.
Given a vertex $u \in V$, its indegree is calculated by the following equation:
\begin{equation}
   \label{eq:indegree}
   \textrm{indegree}(u) = \sum_{v:(v,u)\in{E'}} \min(\sum_{s\in{V_1'},t\in{V_2'}}{x_{vu}^{(s,t)}},1).
\end{equation}



The starting integer programming formulation of fault-tolerance structure generation problem in \cite{xu2017clustered} is shown in Formula~\eqref{equ:ip}.
The objective function in Formula~\eqref{equ:ip} is to minimize the maximum indegree of all the vertices.
The number of binary variables $x_{vu}^{(s,t)}$ is $m \times n \times |E'|$, where $m$ is the number of f-TSVs, $n$ is the number of s-TSVs, while $|E'|$ is the number of edges in split directed graph $G'$.
The constraint (\ref{equ:path}) defines a unit flow from $s$ $\in$ $V_1'$ to $t$ $\in$ $V_2'$, which corresponds a path from $s$, an f-TSV, to $t$, an s-TSV.
The number of this set of constraints is $m\times n\times |V'|$.
The constraint \eqref{equ:nodedj} ensures that a set of $V_2'$ paths, which have the same source $s$ $\in$ $V_1'$, are edge-disjoint. The number of this set of constraints is $m\times (m+n)$.

\begin{figure}[htb!]
\begin{subequations}
\label{equ:ip}
\begin{align}
   \min \ \ & \mathop{\max}\limits_{u\in{V}}\textrm{indegree}(u) \tag{\ref*{equ:ip}}\\
   \textrm{s.t.} \ \
   &\sum_{v:(u,v)\in{E'}}{x_{uv}^{(s,t)}}  -\sum_{v:(v,u)\in{E'}}{x_{vu}^{(s,t)}}= \nonumber \\
   &\quad \left\{
    \begin{array}{ll}
        1,                       & \mbox{if}~ u=s,   \\
        0,                       & \mbox{if}~ u\in{V'-\{s,t\}}, \\
        -1,                      & \mbox{if}~ u=t; \\
    \end{array} \right.                                     \forall s \in V_1', t \in V_2',                     \label{equ:path}\\
    &\sum_{t\in{V_2'}}{x_{uu'}^{(s,t)}} \leq 1,             \quad \forall s\in V_1', (u,u') \in E_1',           \label{equ:nodedj}\\
    &x_{vu}^{(s,t)} \in \{0,1\},                            \quad \forall (v,u) \in E', s \in V_1', t \in V_2'. \label{equ:xuvst}
\end{align}
\end{subequations}
\end{figure}



Though the integer programming method in \cite{xu2017clustered} can generate $K$ fault-tolerance structures using $K$ s-TSVs, the method cannot be directly applied for the generation of adaptive fault-tolerance structures, where the number of s-TSVs might be larger than $K$ in $K$ fault-tolerance structures.
Then a new integer programming formulation is proposed to generate adaptive fault-tolerance structures in minimizing both the used s-TSV number and the multiplexer delay overhead.
The number of s-TSVs used in the structure can be calculated by the Equation~\eqref{eq:usestsv}.
\begin{equation}
   \label{eq:usestsv}
   \textrm{usedstsv} = \sum_{w\in{V_2}} \min(\sum_{s\in{V_1'},t\in{V_2'}}{x_{ww'}^{(s,t)}},1).
\end{equation}

Based on the above notations, the edge-disjoint path search problem can be formulated as the following integer programming \eqref{equ:ip2}.

\begin{figure}[htb!]
\begin{subequations}
\label{equ:ip2}
\begin{align}
   \min \ \ & \{\mathop{\max}\limits_{u\in{V}}\textrm{indegree}(u)+\textrm{usedstsv}\}   \tag{\ref*{equ:ip2}}\\
   \textrm{s.t.} \ \
   &\sum_{v:(u,v)\in{E'}}{x_{uv}^{(s,t)}}  -\sum_{v:(v,u)\in{E'}}{x_{vu}^{(s,t)}}= \nonumber \\
   &\quad \left\{
    \begin{array}{ll}
        v^{(s,t)},               & \mbox{if}~ u=s,   \\
        0,                       & \mbox{if}~ u\in{V'-\{s,t\}}, \\
        -v^{(s,t)},              & \mbox{if}~ u=t; \\
    \end{array} \right.                                     \forall s \in V_1', t \in V_2',         \label{equ:path2}\\
    &\sum_{t\in{V_2'}}{v^{(s,t)}}={K},                      \qquad \forall s\in V_1'.               \label{equ:maxflow}\\
    &v^{(s,t)} \in \{0,1\},                                 \qquad \forall s\in V_1', t \in V_2',   \label{equ:vst}\\
    &\eqref{equ:nodedj}-\eqref{equ:xuvst}.                                                          \notag
\end{align}
\end{subequations}
\end{figure}

Compared with the integer programming~\eqref{equ:ip},
in constraint \eqref{equ:path2} a new binary variable $v^{(s,t)}$ is introduced to indicate whether a unit flow (path) exists from source $s \in V_1'$ to sink $t \in V_2'$.
Besides, a new constraint \eqref{equ:maxflow} is defined to ensure that there will be $K$ paths from each source $s$ $\in$ $V_1'$ to vertices in $V_2'$.
The number of this set of constraints is $m$. By this way, Formula~\eqref{equ:ip2} can be applied for any $K \le n$ and additionally minimize the number of required s-TSVs in the structure,
while Formula~\eqref{equ:ip} can only be applied for the case $K=n$.


Formula \eqref{equ:ip2} is non-linear due to the min-max-min and min-min operations in the objective function. %
Through linearizing the objective function, Formula \eqref{equ:ip2} can be transformed into an integer linear programming (ILP) Formula \eqref{eq:ilp}.
For each edge $(v, u)\in E'$, an extra binary variable $d_{vu}$ and extra constraints \eqref{equ:obj-con-1}-\eqref{equ:obj-con-3} are introduced to replace the min operation in Formula \eqref{eq:indegree} and \eqref{eq:usestsv}. Besides, the extra constraint \eqref{equ:obj-con-4} ensures that the indegrees of all TSVs will not be greater than $\lambda_1$. Another extra constraint \eqref{equ:obj-con-5} ensures that the number of s-TSVs used in the structure equals to $\lambda_2$.

\begin{figure}[htb!]
\begin{subequations}
\label{eq:ilp}
\begin{flalign}
\label{equ:obj-lin}
\ \ \min~~ (\lambda_1+\lambda_2)  && \tag{\ref*{eq:ilp}}
\end{flalign}
\begin{align}
\textrm{s.t.}\ \
    & d_{vu}\geq{x_{vu}^{(s,t)}},                                 && \forall s\in{V_1'}, t\in{V_2'}, (v,u)\in{E'}, \label{equ:obj-con-1}\\
    & d_{vu}\leq{{\sum_{s\in{V_1'},t\in{V_2'}}{x_{vu}^{(s,t)}}}}, && \forall{(v,u)\in{E'}},                        \label{equ:obj-con-2}\\
    & d_{vu} \in \{0,1\},                                         && \forall{(v,u)\in{E'}},                        \label{equ:obj-con-3}\\
    & {\sum_{v:(v,u)\in{E'}}{d_{vu}}}\leq{\lambda_1},             && \forall{u\in{V}},                             \label{equ:obj-con-4}\\
    & {\sum_{(w,w')\in{E_1'}}{d_{ww'}}}={\lambda_2},              && \forall{w\in{V_2}},                           \label{equ:obj-con-5}\\
    & \eqref{equ:nodedj}-\eqref{equ:xuvst}, \eqref{equ:path2}-\eqref{equ:vst}.                                     \notag
\end{align}
\end{subequations}
\end{figure}

%
%
%
%
%
%

For instance, as shown in Fig.~\subref*{fig:splitgraph}, the blue lines present edge-disjoint paths for each split f-TSV,
and the corresponding generated 2 fault-tolerance structure is shown in Fig.~\subref*{fig:special:b}.

%% file: doc/heuristic.tex
\section{Heuristic Framework}
\label{sec:hm}
For large TSV groups, the ILP based method is very time consuming. Consequently,
in this section, we propose a min-cost-max-flow (MCMF) based heuristic method to solve the edge-disjoint path problem. The basic idea is to deal with the f-TSVs one by one and, for each f-TSV, a min-cost-max-flow algorithm is used to find $K$ independent paths.
The edge costs are defined to keep the input port number of multiplexer and the number of s-TSVs as small as possible.

\subsection{Network graph model}
\label{sec:ngm}
In order to find $K$ ($K$ $\leq$ $n$) edge-disjoint paths for an f-TSV $f$ $\in$ $V_1$, we construct a directed graph $G_s(V_s, E_s)$ from $G'$ by adding an extra sink vertex $t$ and some edges.
The vertex set $V_s$ contains two portions, $V_s$ = $V'$ $\cup$ $\{r\}$, and $r$ is the sink vertex.
The edge set $E_s$ = $E'$ $\cup$ $\{V_2' \rightarrow r\}$. 

When finding edge-disjoint paths for a certain TSV $f_i$ $\in$ $V_1$, the edge \textbf{capacities} are defined as follows:
the capacity of the edge from $f_i$ to its splitting vertex $f_i'$ equals to $K$; while the capacities of all the other edges are set to $1$.
The capacity constraints ensure that we can find up to $K$ edge-disjoint paths from $f_i'$ to s-TSV vertices, which correspond to $K$ independent TSV-replacing chains for the TSV $f_i$.



For the splitting edges corresponding to f-TSVs, the edge costs are defined as zero while the splitting edges of s-TSVs are defined as follows.
\begin{equation}
   ec_s(w,w') = \left\{
   \begin{array}{ll}
	0,  \textrm{ if } (w,w') \in E_1', w \in V_2, \textrm{ and $w$ has}\\
    \textrm{ been used}.\\
	C^{K}, \textrm{ if }(w,w') \in E_1', w \in V_2, \textrm{ and $w$}\\
    \textrm{ has not been used}.
   \end{array} \right.
   \label{Eq.9}
\end{equation}

$C$ is constant, which represents the costs of introducing a new s-TSV for constructing the fault-tolerance structure. And the edge costs tend to restrict the use of s-TSVs.
In the experiment, we set $C$ to 3 by the experimental results shown in Section~\ref{sec:eftsg}.

For the edges in $E_2'$, which correspond to the replaceable relations between TSVs, the edge costs are defined as follows.
\begin{equation}
   ec_s(u,v) = \left\{
   \begin{array}{ll}
     0,  \textrm{ if } (u,v) \in E_2'\textrm{ and $(u,v)$ corresponds}\\
       \textrm{ to a TSV connection} \\
     C^{tc[v]}, \textrm{ if } (u,v) \in E_2' \textrm{ and $(u,v)$ does not}\\
	   \textrm{ correspond to a TSV connection}\\
   \end{array} \right.
   \label{Eq.10}
\end{equation}

In the edge cost function \eqref{Eq.10}, $tc[v]$ is defined to be the number of edges that end at $v$ and have been used as TSV connections in the generated partial fault-tolerance structure, that is, the edges that have been traversed by edge-disjoint paths of some other f-TSVs. Therefore, $tc[v]$ corresponds to the input port number of the multiplexer in the input side of the TSV $v$.

With this edge costs function, firstly, we tend to make full use of existing TSV connections to build the edge-disjoint paths for the current f-TSV since it will not increase the input ports of the multiplexers.

Secondly, to minimize the maximum size of multiplexers, the costs of the edges that do not correspond to TSV connections are defined as the exponential function of $tc[v]$.

\begin{algorithm}[tb!]
 \textbf{Input}: A directed graph $G'(V',E')$, which contains $m$ f-TSVs and $n$ s-TSVs.\\
 \textbf{Output}: A repairable structure including $m$ $\times$ $K$ paths.
 \begin{algorithmic}[1]
  \For{f-TSV $f_i \gets 1$ to $m$}
      \State Construct a directed graph $G_s(V_s, E_s)$ for $f_i$;
      \State \Comment{Find $K$ edge-disjoint paths for $f_i$;}
      \State Solve the MCMF model for $f_i$;
  \EndFor
  \State \Comment{Perturb the repairable structure;}
  \While{no coverage}
      \State Randomly select an f-TSV $f_i$;
      \State Resolve edge-disjoint paths for $f_i$ by MCMF;
      \State Record the maximum number of TSV connections on all TSVs;
  \EndWhile
 \end{algorithmic}
  \caption{\textbf{Pseudo code of our heuristic method}}
  \label{alg1}
\end{algorithm}

\begin{figure*}[tb!]
    \centering
    \hspace{-.1in}
    \subfloat[]{\includegraphics[width=.32\textwidth]{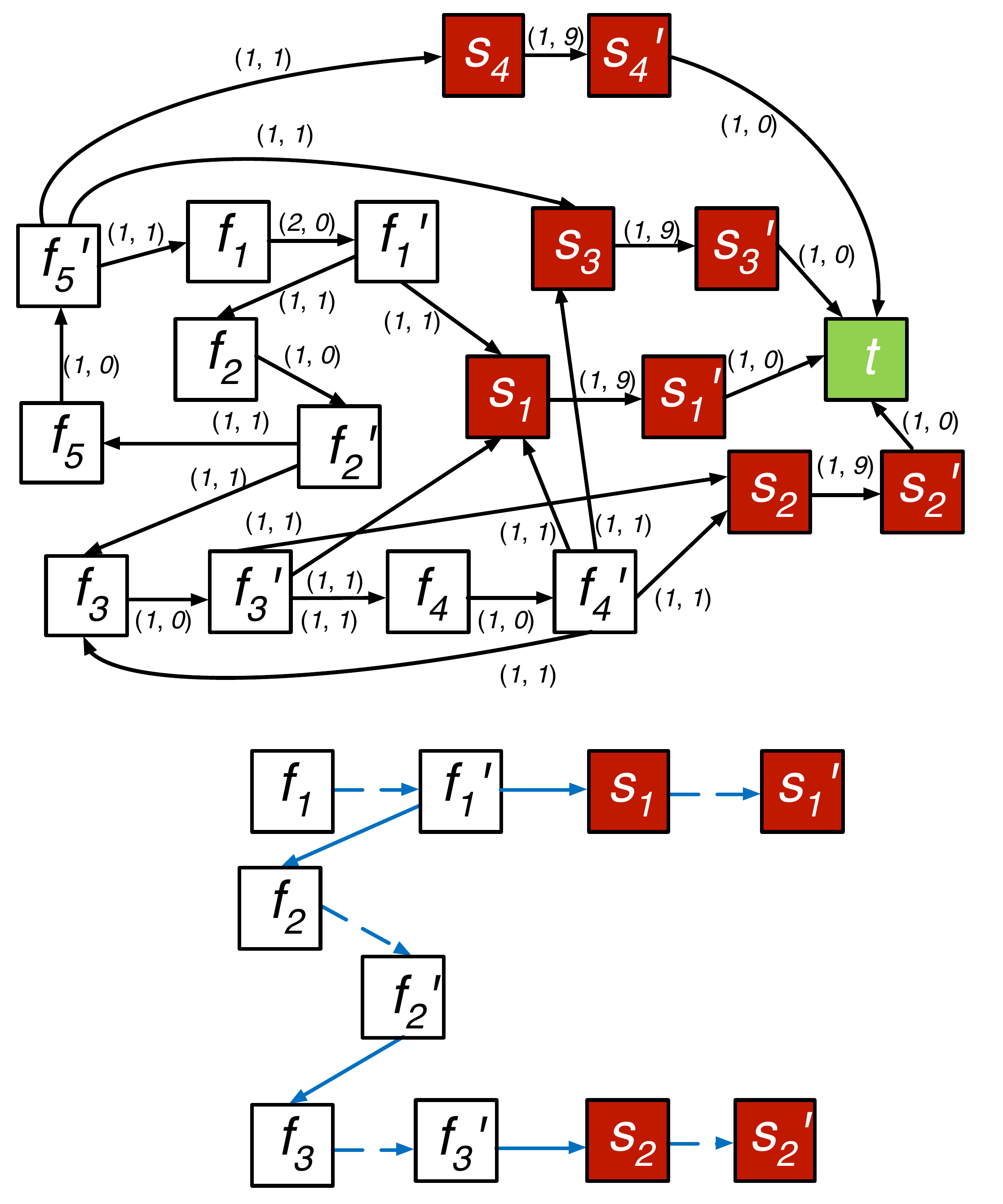} \label{fig:mcmfgraph-1}}
    \hspace{.05in}
    \subfloat[]{\includegraphics[width=.32\textwidth]{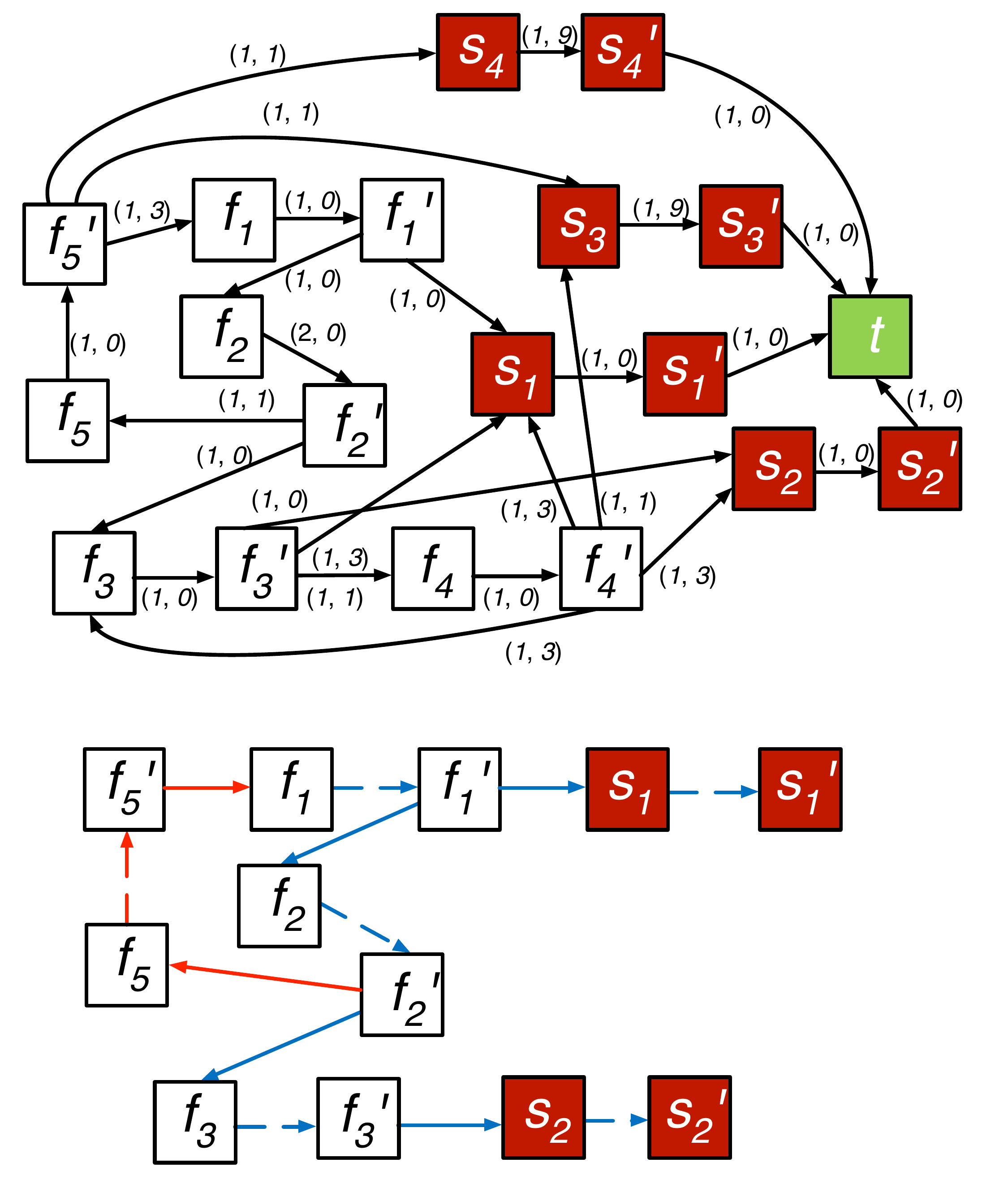} \label{fig:mcmfgraph-2}}
    \hspace{.05in}
    \subfloat[]{\includegraphics[width=.32\textwidth]{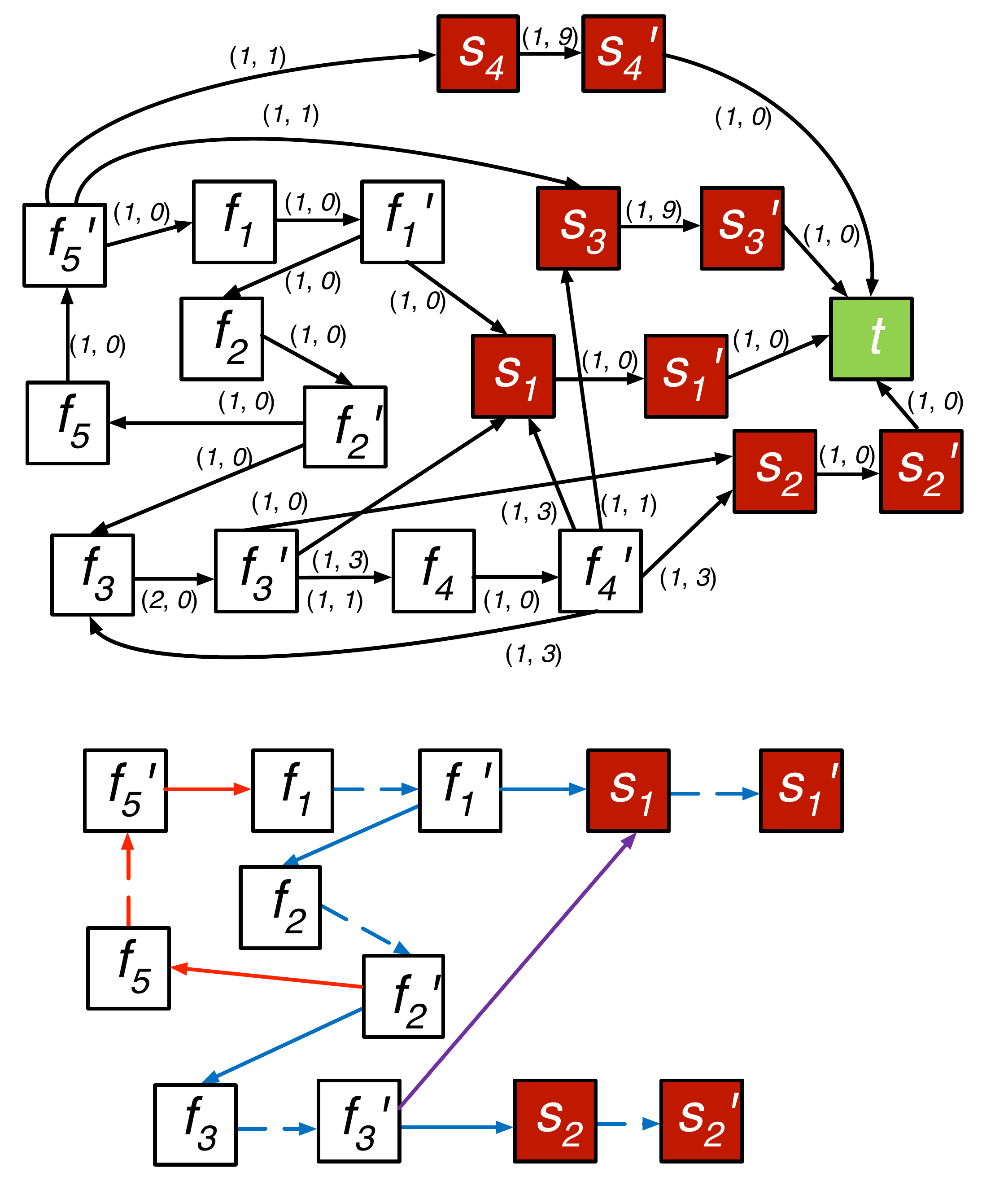} \label{fig:mcmfgraph-3}}
    \hspace{-.1in}
    \caption{\revise{
    Label on edges represents ($capacity$, $cost$):
    (a) The min-cost-max-flow network for f-TSV $f_1'$, where the two edge-disjoint paths for $f_1'$: \{$f_1'$ $\rightarrow$ $s_1$ $\rightarrow$ $s_1'$\} and \{$f_1'$ $\rightarrow$ $f_2$ $\rightarrow$ $f_2'$ $\rightarrow$ $f_3$ $\rightarrow$ $f_3'$ $\rightarrow$ $s_2$ $\rightarrow$ $s_2'$\};
    (b) After solving $f_1'$, the min-cost-max-flow network for f-TSV $f_2'$, where the two edge-disjoint paths for $f_2'$:
    \{$f_2'$ $\rightarrow$ $f_5$ $\rightarrow$ $f_5'$ $\rightarrow$ $f_1$ $\rightarrow$ $f_1'$ $\rightarrow$ $s_1$ $\rightarrow$ $s_1'$\} and \{$f_2'$ $\rightarrow$ $f_3$ $\rightarrow$ $f_3'$ $\rightarrow$ $s_2$ $\rightarrow$ $s_2'$\};
    (c) After solving $f_1'$ and $f_2'$, the min-cost-max-flow network for f-TSV $f_3'$, where the two edge-disjoint paths for $f_3'$:
    \{$f_3'$ $\rightarrow$ $s_1$ $\rightarrow$ $s_1'$\} and \{$f_3'$ $\rightarrow$ $s_2$ $\rightarrow$ $s_2'$\}.
    }}
    \label{fig:mcmfgraph}
\end{figure*}

\begin{figure}[tb!]
    \centering
    \includegraphics[width=.32\textwidth]{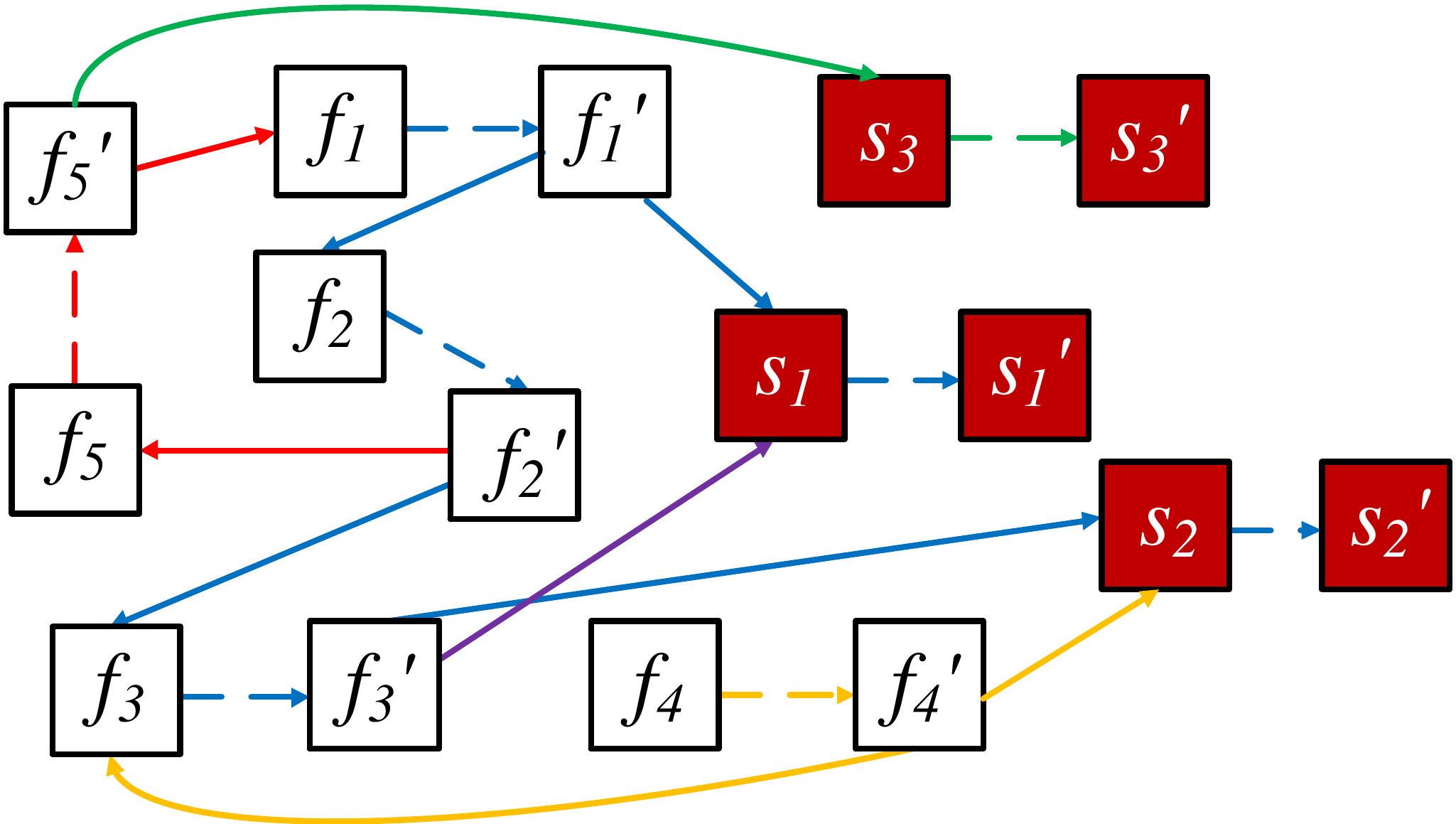}
    \caption{The generated 2-fault tolerance structure \revise{by solving edge-disjoint paths for all f-TSVs}, where the TSV connections are shown in solid edges.}
    \label{fig:f4s}
\end{figure}

\subsection{Algorithmic flow of heuristic}
The algorithmic flow of the proposed heuristic is summarized in Algorithm \ref{alg1}.
Because the quality of solution depends on the order of f-TSVs selected, an iterative post-processing stage is used to improve the generated fault-tolerance structures.
In the post-processing stage, we randomly select an f-TSV\revise{, and} define the edge costs based on the TSV paths of all the other f-TSVs.
\revise{Then} we re-solve the min-cost-max-flow model to find edge-disjoint paths for the selected f-TSV.
The procedure is repeated until \revise{the multiplexer maximum input port number keeps} unchanged over a predefined threshold \revise{iteration} number.

Fig.~\subref*{fig:mcmfgraph-1} -- Fig.~\subref*{fig:mcmfgraph-3} illustrate the process of the heuristic method.
We choose the f-TSV $f_1'$ to start with. The min-cost-max-flow network for $f_1'$ is shown in Fig.~\subref*{fig:mcmfgraph-1}.
All the costs of edges that end at f-TSVs and s-TSVs are initialized at $1$ since there are no any other f-TSV paths and for all $v$, $tc[v]=0$.
By solving the min-cost-max-flow, $2$ edge-disjoint paths, which correspond to two independent TSV replacing chains for $f_1$, are obtained and the TSV connections (solid edges) in the partial fault-tolerance structure.

With the $2$ edge-disjoint paths for $f_1$, the flow network is updated (edge costs and capacities) for f-TSV $f_2'$ and shown in Fig.~\subref*{fig:mcmfgraph-2}.
The edges that are on the edge-disjoint paths of $f_1$ have zero costs.
Considering the vertex $s_1$, for example, the edge $(f_1',s_1)$ has zero costs since it has been traversed by the TSV path of $f_1$ while the edges $(f_3',s_1)$ and $(f_4',s_1)$ have a cost of $3$ because the both edges are not traversed by any TSV paths of $f_1$ and $tc[s_1] = 1$.
A new TSV connection will be introduced if we use $(f_3',s_1)$ or $(f_4',s_1)$ on the edge-disjoint paths for $f_2$, which increase the input ports of multiplexer in the input side of the TSV $s_1$.
With the updated network, we can find two edge-disjoint paths from $f_2'$ to s-TSVs by making use of the existing TSV connections as many as possible, which potentially reduces the TSV connections on individual TSVs and minimizes the maximum number of the input ports of multiplexers.
The bottom part of Fig.~\subref*{fig:mcmfgraph-2} shows the TSV connections in the updated partial fault-tolerance structure.

Repeating the same process until the min-cost-max-flow model is solved for all f-TSVs, we obtain $2$ edge-disjoint paths from each split f-TSV vertex in $V_1'$, $f_1'\cdots f_5'$, to split s-TSV vertices in $V_2'$, $s_1' \cdots s_3'$,
\revise{as shown in Fig.~\ref{fig:f4s}. Here the solid edges are TSV connections.}

%% file: doc/flow.tex
\section{Fault Tolerance TSV Planning}
\label{sec:mgy}

In this section, we discuss a top-down fault tolerance TSV planning framework to form f-TSV groups and generate adaptive fault-tolerance structures.
The number of f-TSV groups is greatly reduced as well as the total number of s-TSVs because of adaptive fault-tolerance structures.

Given an f-TSV planning result and the floorplan of the blocks, we know the number and positions of all f-TSVs.
Then f-TSV groups are firstly formed using a top-down iterative f-TSV partitioning under the yield constraint and, then, the adaptive fault-tolerance structures are generated for each group.
In each iteration of the f-TSV partitioning stage, the group with the smallest yield will be partitioned into two new f-TSV groups using the min-cut bi-partitioning algorithm and the required s-TSVs are also allocated for evaluating the group yield.
The iterative f-TSVs partitioning is repeated until the target chip yield is satisfied.
Therefore, the number and position of required s-TSVs for each f-TSV group are determined simultaneously in the f-TSV partitioning stage.

The chip yield is the product of group yield, which depends on the maximum number of tolerant faults ($K$), the number of TSVs, and the defect probability of TSVs as discussed in Section~\ref{subsec:cty}.
We construct the replaceable relation graph $G$, whose vertex set includes the f-TSVs in the group and the corresponding candidate s-TSVs, for computing $K$ and allocating s-TSVs.
The maximum number of tolerant faults, $K$, can be determined in polynomial time by solving a max-flow problem on $G$, as discussed in Section~\ref{sec:ta}.
The min-cost-max-flow based heuristic in Section~\ref{sec:hm} is used to temporarily generate an adaptive $K$-fault tolerance structure, thus the number of required s-TSVs are determined.

\begin{figure}[tb!]
  \centering
  \includegraphics[width=.50\textwidth]{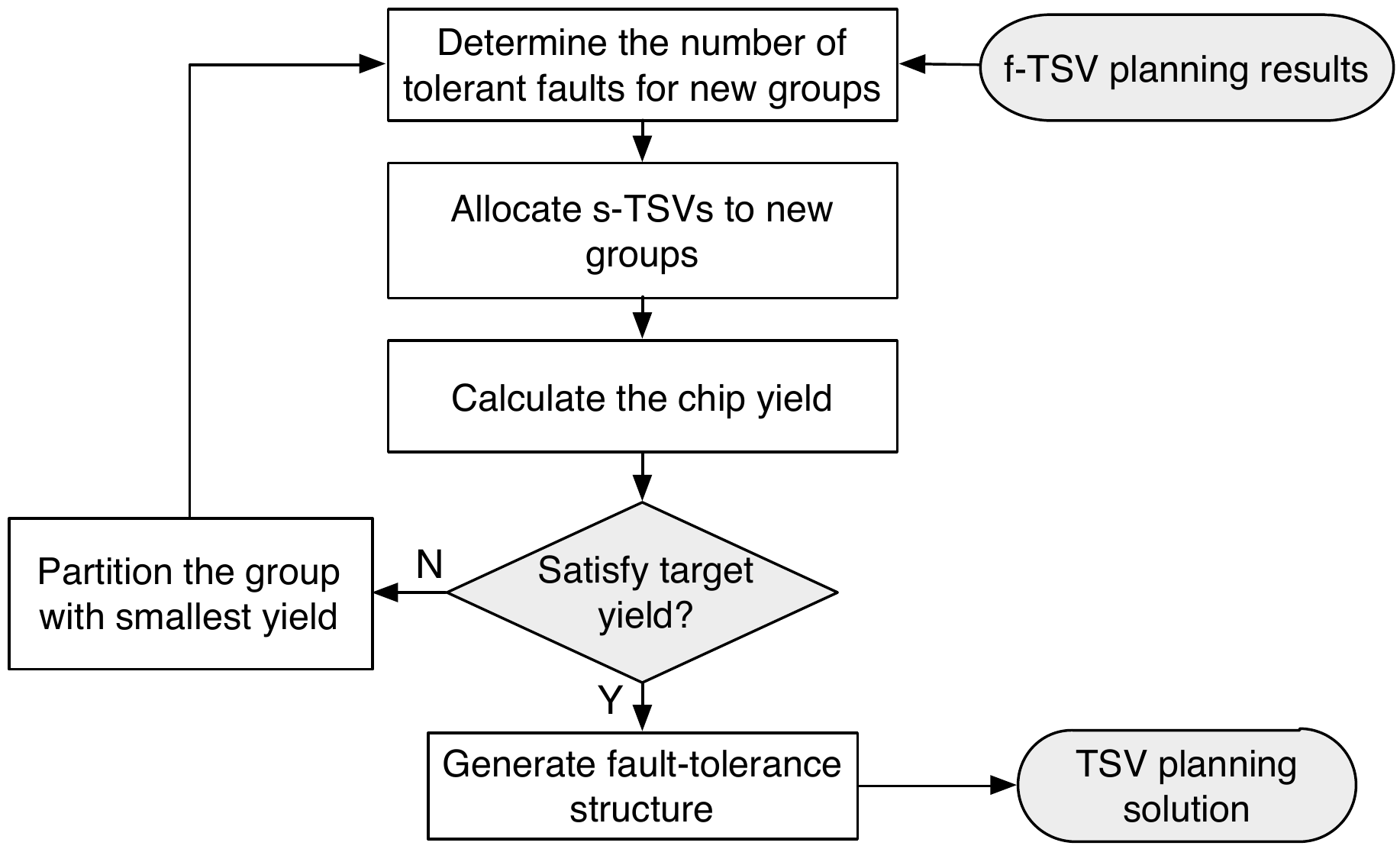}
  \caption{The flow of the proposed fault tolerance TSV planning.}
  \label{fig:flow}
\end{figure}

Finally, the ILP based method in Section~\ref{sec:ilp} and the min-cost-max-flow (MCMF) based heuristic in Section~\ref{sec:hm} can be adopted to generate adaptive fault-tolerance structures with minimization of both the multiplexer delay overhead and the hardware cost.
Fig.~\ref{fig:flow} illustrates the proposed TSV planning framework.

In~\cite{wang2015defect}, a greedy method is used to partition f-TSVs into groups and then an ILP formulation is adopted to allocate s-TSVs for each group. The generation of fault-tolerance structure is not considered since they assume regular structures always exist.
In~\cite{xu2017clustered}, the TSV planning framework includes a top-down partitioning followed by a bottom-up iterative merging (clustering) for reducing the number of f-TSV groups. Then, a min-cost-max-flow based method is used to allocate s-TSVs for each group and an ILP model is adopted to generate fault-tolerance structures.
The same number of s-TSVs are allocated to all the f-TSV groups in \cite{wang2015defect,xu2017clustered} and, for an f-TSV group, the key point is to ensure enough number of candidate s-TSVs that can be shared by all the f-TSVs in the group. As a result, many small f-TSV groups are formed, which potentially causes an overuse of s-TSVs. 

Compared with the above mentioned two works, the proposed TSV planning framework includes a similar top-down partitioning stage, but the allocation of s-TSVs during the partitioning is very different. That is because adaptive fault-tolerance structures with various number of s-TSVs are built temporarily by solving a sequence of min-cost max-flow problem. 

%% file: doc/result.tex
\begin{table*}[tbp]
\centering \caption{Comparison between ILP \cite{xu2017clustered} and our methods for generating adaptive fault-tolerance structure.}
\label{tab:tab2}
\renewcommand{\arraystretch}{1.2}
\resizebox{18.0cm}{!} {
\begin{tabular}{|c|c|c|c|c|c|c|c|c|c|c|c|c|c|c|c|c|}
    \hline
    \multirow{2}{*}{Graph}  &\multirow{2}{*}{$m$} &\multirow{2}{*}{$n$} &\multirow{2}{*}{\#Edges} &\multirow{2}{*}{$K$} &\multicolumn{4}{c|}{ILP \cite{xu2017clustered}} &\multicolumn{4}{c|}{ILP} &\multicolumn{4}{c|}{Heuristic} \\
    \cline{6-17}
    & & & & & \#Port & \#us &\tabincell{c}{IWire(um)\\(ratio)} & RT(s) & \#Port & \#us & \tabincell{c}{IWire(um)\\(ratio)} & RT(s) & \#Port & \#us & \tabincell{c}{IWire(um)\\(ratio)} & RT(s)\\
    \hline \hline
    \texttt{$G_{11}$} & 9 & 4 & 72 & 3 & 3 & 3 & 32.90 (0.51\%) & 535.20 & 3 & 3 & 32.90 (0.51\%)& 301.53 & 3 & 4 & 25.88 (0.40\%)& 0.008\\
    \texttt{$G_{12}$} & 13 & 4 & 129 & 2 & 3 & 2 & 6.85 (0.18\%) & 603.68 & 3 & 2 & 9.65 (0.25\%) & 67.80 & 3 & 4 & 16.79 (0.43\%)& 0.013\\
    \texttt{$G_{13}$} & 14 & 4 & 101 & 1 & NA & NA & NA & $>$3600 & 2 & 4  & 29.77 (1.77\%) & 1.09 & 3 & 4 & 28.99 (1.72\%) & 0.006\\
    \texttt{$G_{14}$} & 15 & 5 & 177 & 2 & NA & NA & NA & $>$3600 & 3 & 4 & 32.50 (0.62\%) & 96.90 & 3 & 4 & 32.71 (0.62\%) & 0.009\\
    \texttt{$G_{15}$} & 18 & 5 & 215 & 2 & NA & NA & NA & $>$3600 & 3 & 4 & 65.20 (0.96\%) & 240.07 & 4 & 5 & 52.35 (0.77\%) & 0.013\\
    \texttt{$G_{16}$} & 18 & 6 & 199 & 2 & NA & NA & NA & $>$3600 & 3 & 6 & 90.03 (1.76\%) & 155.74 & 3 & 6 & 98.36 (1.93\%) & 0.011\\
    \texttt{$G_{17}$} & 21 & 7 & 255 & 2 & NA & NA & NA & $>$3600 & NA & NA & NA & $>$3600 & 4 & 6 & 214.39 (1.83\%) & 0.017\\
    \texttt{$G_{18}$} & 26 & 13 & 529 & 4 & NA & NA & NA & $>$3600 & NA & NA & NA & $>$3600 & 4 & 12 & 333.60 (1.47\%) & 0.038\\
    \hline
    \texttt{$G_{21}$} & 9 & 5 & 99 & 5 & 4 & 5 & 16.34 (0.15\%) & 100.84 & 4 & 5 & 16.34 (0.15\%) & 101.10 & 4 & 5 & 16.34 (0.15\%) & 0.005\\
    \texttt{$G_{22}$} & 12 & 5 & 155 & 5 & 5 & 5 & 49.77 (0.25\%) & 304.91 & 5 & 5 & 49.77 (0.25\%) & 306.14 & 6 & 5 & 56.26 (0.28\%) & 0.007\\
    \texttt{$G_{23}$} & 14 & 5 & 197 & 5 & 5 & 5 & 10.21 (0.06\%) & 3435.64 & 5 & 5 & 10.21 (0.06\%) & 3468.93 & 5 & 5 & 11.84 (0.07\%) & 0.010\\
    \texttt{$G_{24}$} & 16 & 5 & 225 & 5 & 5 & 5 & 108.19 (0.71\%) & 3519.16 & 5 & 5 & 108.19 (0.71\%) & 3519.16 & 7 & 5 & 123.18 (0.81\%) & 0.016\\
    \texttt{$G_{25}$} & 18 & 5 & 329 & 5 & NA & NA & NA & $>$3600 & NA & NA & NA & $>$3600 & 5 & 5 & 72.01 (0.26\%) & 0.016\\
    \texttt{$G_{26}$} & 23 & 6 & 467 & 6 & NA & NA & NA & $>$3600 & NA & NA & NA & $>$3600 & 6 & 6 & 45.99 (0.12\%) & 0.027\\
    \texttt{$G_{27}$} & 24 & 6 & 550 & 6 & NA & NA & NA & $>$3600 & NA & NA & NA & $>$3600 & 6 & 6 & 30.65 (0.08\%) & 0.034\\
    \texttt{$G_{28}$} & 25 & 7 & 524 & 7 & NA & NA & NA & $>$3600 & NA & NA & NA & $>$3600 & 7 & 7 & 24.65 (0.06\%) & 0.037\\
    \hline
\end{tabular}
}
\end{table*}

\section{Experimental Results}
\label{sec:result}
The proposed algorithms have been implemented in C++ language and tested on a 12-core 2.0 GHz Linux server with 64 GB RAM.
The TSV pitch is assumed to be 5\emph{um}$\times$5\emph{um} \cite{ITRS}.
{LEDA} \cite{TOOL_B1999_leda} is adopted to solve the max-flow and the min-cost-max-flow problems.
{GLPK} \cite{TOOL_glpk} is used as the ILP solver.
{hMetis} \cite{karypis1999multilevel} is adopted on f-TSVs partitioning.

\subsection{Effectiveness and Efficiency of Fault-Tolerance Structure Generation Method}
\label{sec:eftsg}

%

We generate several TSV replaceable relation graphs $G_{11}$--$G_{18}$ by using the proposed TSV planning framework on MCNC and GSRC benchmarks.
Each graph contains f-TSVs and the corresponding candidate s-TSVs, which are covered by at least one of the bounding boxes of the f-TSVs.
In order to compare the proposed ILP model with the ILP method in \cite{xu2017clustered} on $G_{11}$--$G_{18}$, we adapt the ILP formulation in \cite{xu2017clustered} here.
To generate the $K$-fault tolerance structure on a TSV replaceable relation graph $G$, we select $K$ s-TSVs in all $n$ s-TSVs, and unit flow constraints are defined from all f-TSVs to those chosen $K$ s-TSVs.
If the $K$-fault tolerance structure is still not achieved after solving all $K$ combinations, we think the ILP method in \cite{xu2017clustered} cannot generate  the $K$-fault tolerance structure on this TSV replaceable relation graph $G$.

In addition, the previous \revise{work} in~\cite{xu2017clustered} \revise{deals} with a special type of TSV fault-tolerance structure generation.
That is, they are under an assumption that a predetermined number of s-TSVs are assigned to each TSV group, and an f-TSV in a group should be replaced by any s-TSV within the group.
We also generate some specific TSV replaceable relation graphs $G_{21}$--$G_{28}$ by using the TSV planning methods in \cite{xu2017clustered} on MCNC and GSRC benchmarks.
Since the f-TSVs can be replaced by all $n$ s-TSVs in each graph, the $n$-fault tolerance structure always exists.

First, we show the effectiveness of the proposed ILP model. TABLE~\ref{tab:tab2} shows the experimental results, where ``ILP" and ``Heuristic'' denote results of the proposed ILP model and min-cost-max-flow based heuristic method, respectively.
Columns ``$m$'', ``$n$'', ``\#Edges'', and ``$K$'' list the number of f-TSVs, the total number of available s-TSVs, the number of edges, and the number of maximumly tolerant faults on each TSV replaceable relation graph.
Besides, \revise{columns} ``\#Port'' and ``\#us'' show the maximum port number of multiplexers and the number of s-TSVs used in the generated fault-tolerance structure.
``IWire'' shows the sum of incremental half-perimeter wirelength overhead of all f-TSVs incurred by the fault-tolerance structure, and the ratio of ``IWire'' to the sum of net wirelength of all f-TSVs is listed in ``ratio''.
``RT'' reports the total computational time in seconds.
``NA'' represents  that the $K$-fault tolerance structure cannot be achieved within the time limit (3600$s$).
As shown in TABLE~\ref{tab:tab2}, the ILP method in \cite{xu2017clustered} generates the fault-tolerance structure only on two smallest graphs.
However, the proposed ILP formulation can achieve the fault-tolerance structure on six graphs.

Second, we show the efficiency of the proposed heuristic method.
TABLE~\ref{tab:tab2} also compares the proposed heuristic method with the proposed ILP method.
It can be noticed that, on small graphs $G_{11}$--$G_{16}$ and $G_{21}$--$G_{24}$, the fault-tolerance structure generated by ILP has smaller maximum port number of multiplexers and used less s-TSV numbers than that generated by the heuristic method.
Therefore, for small TSV replaceable relation graphs, ILP can achieve an optimal solution, which can be used to verify the accuracy of the solution of the heuristic method. But since ILP is an NP-hard problem, its runtime increases dramatically with the size of TSV replaceable relation graphs.
As shown in TABLE~\ref{tab:tab2}, the ILP method cannot generate the fault-tolerance structure on large graphs $G_{17}$--$G_{18}$ and $G_{25}$--$G_{28}$ within the time limit (3600s). Therefore, for large TSV replaceable relation graphs, the ILP based method is very time consuming, which can indirectly demonstrate the efficiency of the proposed heuristic method.

\begin{table}[!tb]
\renewcommand{\arraystretch}{1.00}
\centering \caption{Effect of $C$ on s-TSV numbers and maximum port number of multiplexers.}
\label{tab:tab9}
\resizebox{6.6cm}{!} {
\begin{tabular}{|c|c|c|c|c|}
    \hline
    \multirow{2}{*}{Benchmark}  &\multicolumn{2}{c|}{$C$ = 2}  &\multicolumn{2}{c|}{$C$ = 3} \\
    \cline{2-5}
    &\#s-TSV  & \#Port & \#s-TSV  & \#Port \\
    \hline \hline
    \texttt{ami33} & 52  & 4 & 46 & 4\\
    \texttt{ami49} & 80  & 8 & 66 & 6\\
    \texttt{n50}   & 108 & 7 & 98 & 7\\
    \texttt{n100}  & 181 & 8 & 169 & 7\\
    \texttt{n200}  & 267 & 7 & 250 & 7 \\
    \texttt{n300}  & 395 & 8 & 381 & 6\\
    \hline
\end{tabular}
}
\end{table}

\begin{table*}[tbp!]
\centering \caption{Comparisons among \cite{wang2015defect}, \cite{xu2017clustered}, and the proposed adaptive fault-tolerance structure (AFTS) under $3$-fault tolerance structures (target yield = 99.7\%, $p$ = 0.001).}
\label{tab:tab7}
\renewcommand{\arraystretch}{1.2}
\resizebox{18.3cm}{!} {
\begin{tabular}{|c|c|c|c|c|c|c|c|c|c|c|c|c|c|c|c|c|c|c|c|}
    \hline
    \multirow{2}{*}{Bench}   &\multirow{2}{*}{\#f-TSV}   &\multicolumn{3}{c|}{\cite{wang2015defect}} &\multicolumn{5}{c|}{\cite{xu2017clustered}} & \multicolumn{5}{c|}{AFTS (K$\leq$3)} & \multicolumn{5}{c|}{AFTS (maximum $K$)}\\
    \cline{3-20}
    &&\#s-TSV & \#gp & Yield & \#s-TSV & \#gp & \#Port & K & Yield & \#s-TSV & \#gp & \#Port & K & Yield & \#s-TSV & \#gp & \#Port & K & Yield\\
    \hline \hline
    \texttt{ami33} & 55 & 48 & 16 & 100\% & 48 & 16 & 4 & 3 &  100\% & 31 & 2 & 3 & 3 & 100\% & 46 & 2 & 4 & 4 & 100\% \\
    \texttt{ami49} & 130 & 72 & 24 & 100\% & 66 & 22 & 5 & 3 & 100\% & 54  & 2 & 5 & 3 & 99.99\% & 66 & 2 & 6 & 5 & 100\% \\
    \texttt{n50}   & 386 & 210 & 70 & 99.97\% & 204 & 68 & 7 & 3 & 100\% & 82  & 5 & 6 & 2 & 99.96\% & 98 & 5 & 7 & 5 & 99.98\% \\
    \texttt{n100}  & 592 & 294 & 98 & 99.91\% & 291 & 97 & 7 & 3 & 99.94\% & 136  & 7 & 6 & 3 & 99.91\% & 169 & 7 & 7 & 6 & 99.93\% \\
    \texttt{n200}  & 1127 & 396 & 132 & 99.86\% & 393 & 131 & 6 & 3 & 99.86\% & 179  & 8 & 5 & 3 & 99.85\% & 250 & 8 & 7 & 6 & 99.86\% \\
    \texttt{n300}  & 1232 & 501 & 167 & 99.81\% & 498 & 166 & 6 & 3 & 99.83\% & 246  & 9 & 5 & 3 & 99.78\% & 381 & 7 & 6 & 6 & 99.80\% \\
    \texttt{t337}&640&315&105&99.90\%&309&103&4&3        &99.91\%&158&8&5&3&99.88\%&214&6&6&6         &99.90\%\\
    \texttt{t469}&1546&600&200&99.71\%&588&196&6&3        &99.73\%&313&11&6&3&99.71\%&412&9&7&7         &99.72\%\\
    \hline
    avg. &714&305&102&99.90\%&300&100&6&3
    &99.91\%&150&7&5&3&99.89\%&205&6&7&6 &99.90\%\\
    ratio & -- & +32.79\% & --  & -- & +31.67\% & -- &  -- & -- & -- & -26.83\% & -- & -- & --  & -- & 1.00 & -- & -- & -- & -- \\
    \hline
\end{tabular}
}
\end{table*}

In addition, the parameter $C$ in edge cost \revise{functions}~\eqref{Eq.9} and~\eqref{Eq.10} is also set through experimental results.
The experiment is performed on MCNC and GSRC benchmarks.
In the experiment, if $C$ is set to 4, some edge cost values are out of bound, which cannot be solved by min-cost-max-flow based model.
And we also set $C$ to 2 and 3, the number of used s-TSVs and maximum port number of multiplexers varied with $C$, which is shown in TABLE~\ref{tab:tab9}.
\revise{Columns} ``\#s-TSV'' and ``\#Port'' list the total number of allocated s-TSVs and the maximum port number of multiplexers among all f-TSV groups.
We noticed that compared with $C$ = 2, $C$ = 3 can achieve a fault tolerance structure with less number of used s-TSVs and smaller maximum port number of multiplexers.
Therefore, in the experiment, we set $C$ to 3.

\begin{table*}[tbp!]
\centering \caption{Comparisons among \cite{wang2015defect}, \cite{xu2017clustered}, and the proposed adaptive fault-tolerance structure (AFTS) under $3$-fault tolerance structures (target yield = 99.5\%, $p$ = 0.01).}
\label{tab:tab10}

\renewcommand{\arraystretch}{1.2}
\resizebox{18.3cm}{!} {
\begin{tabular}{|c|c|c|c|c|c|c|c|c|c|c|c|c|c|c|c|c|c|c|c|}
    \hline
    \multirow{2}{*}{Bench}   &\multirow{2}{*}{\#f-TSV}   &\multicolumn{3}{c|}{\cite{wang2015defect}} &\multicolumn{5}{c|}{\cite{xu2017clustered}} & \multicolumn{5}{c|}{AFTS (K$\leq$3)} & \multicolumn{5}{c|}{AFTS (maximum $K$)}\\
    \cline{3-20}
    &&\#s-TSV & \#gp & Yield & \#s-TSV & \#gp & \#Port & K & Yield & \#s-TSV & \#gp & \#Port & K & Yield & \#s-TSV & \#gp & \#Port & K & Yield\\
    \hline \hline
    \texttt{ami33} & 54 & 51 & 17 & 100\% & 51 & 17 & 4 & 3 &  100\% & 35 & 4 & 3 & 3 & 100\% & 48 & 4 & 4 & 4 & 100\% \\
    \texttt{ami49} & 130 & 87 & 29 & 99.96\% & 81 & 27 & 5 & 3 & 99.96\% & 62 & 5 & 4 & 3 & 99.94\% & 73 & 5 & 5 & 4 & 99.95\% \\
    \texttt{n50}   & 388 & 231 & 77 & 99.89\% & 222 & 74 & 6 & 3 & 99.92\% & 102  & 8 & 5 & 3 & 99.88\% & 113 & 8 & 7 & 5 & 99.90\% \\
    \texttt{n100}  & 589 & 330 & 110 & 99.84\% & 324 & 108 & 6 & 3 & 99.87\% & 165  & 12 & 5 & 3 & 99.84\% & 194 & 11 & 7 & 6 & 99.87\% \\
    \texttt{n200}  & 1130 & 438 & 146 & 99.73\% & 435 & 145 & 7 & 3 & 99.74\% & 210  & 17 & 6 & 2 & 99.72\% & 280 & 15 & 7 & 6 & 99.73\% \\
    \texttt{n300}  & 1236 & 555 & 185 & 99.62\% & 549 & 183 & 6 & 3 & 99.63\% & 295  & 20 & 5 & 3 & 99.60\% & 426 & 20 & 6 & 5 & 99.61\% \\
    \texttt{t337}  & 637 & 342 & 114 & 99.82\% & 330 & 110 & 4 & 3  & 99.82\% & 184 & 13 & 4 & 3 & 99.78\% & 227 & 12 & 7 & 7 & 99.81\%\\
    \texttt{t469}  & 1553 & 645 & 215 & 99.55\% & 633 & 211 & 7 & 3 & 99.56\% & 352 & 25 & 6 & 3 & 99.52\% & 455 & 23 & 7 & 6 & 99.55\%\\
    \hline
    avg. & 715 & 335 & 112 & 99.80\% & 329 & 110 & 6 & 3 & 99.81\% & 176 & 13 & 5 & 3 & 99.79\% & 227 & 12 & 7 & 6 & 99.80\%\\
    ratio & -- & +32.24\% & --  & -- & +31.01\% & -- &  -- & -- & -- & -22.47\% & -- & -- & --  & -- & 1.00 & -- & -- & -- & -- \\
    \hline
\end{tabular}
}
\end{table*}

\begin{table*}[tbp!]
\centering \caption{Comparisons among \cite{chen2015novel}, \cite{wang2015defect}, \cite{xu2017clustered}, and the proposed adaptive fault-tolerance structure (AFTS) under $1$-fault tolerance structures (target yield = 99.5\%).}
\label{tab:tab8}
\renewcommand{\arraystretch}{1.2}
\resizebox{18.0cm}{!} {
\begin{tabular}{|c|c|c|c|c|c|c|c|c|c|c|c|c|c|c|c|c|}
    \hline
    \multirow{2}{*}{Bench}   &\multirow{2}{*}{\#f-TSV}   &\multicolumn{3}{c|}{\cite{wang2015defect}} &\multicolumn{4}{c|}{\cite{chen2015novel}} & \multicolumn{4}{c|}{\cite{xu2017clustered}} & \multicolumn{4}{c|}{AFTS ($K$=1)}\\
    \cline{3-17}
    &&\#s-TSV & \#gp & Yield & \#s-TSV & \#gp & \#Port & Yield & \#s-TSV & \#gp & \#Port & Yield & \#s-TSV & \#gp & \#Port & Yield\\
    \hline \hline
    \texttt{ami33} & 52 & 16 & 16 & 99.99\% & 16 & 16 & 4 & 99.99\% & 16 & 16 & 3 &  99.99\%  & 13 & 2 & 2 & 99.99\% \\
    \texttt{ami49} & 124 & 28 & 28 & 99.95\% & 25  & 25 & 5 & 99.96\% & 25 & 25 & 4 & 99.96\% & 22 & 3 & 3 & 99.95\% \\
    \texttt{n50}   & 383 & 74 & 74 & 99.84\% & 68  & 68 & 8 & 99.87\% & 68 & 68 & 4 & 99.87\% & 53 & 8 & 3 & 99.84\% \\
    \texttt{n100}  & 596 & 108 & 108 & 99.65\% & 95 & 95 & 8 & 99.68\% & 95 & 95 & 5 & 99.68\% & 78 & 12 & 4 & 99.64\% \\
    \texttt{n200}  & 1126 & 141 & 141 & 99.61\% & 132  & 132 & 8 & 99.64\% & 132 & 132 & 6 & 99.64\% & 110 & 22 & 5 & 99.61\% \\
    \texttt{n300}  & 1230 & 197 & 197 & 99.51\% & 183  & 183 & 9 & 99.53\% & 183 & 183 & 6 & 99.53\% & 158 & 31 & 5 & 99.51\% \\
    \texttt{t337} &639&124&124&99.65\%&113&113&8
    &99.67\%&113&113&6&99.67\%&91&16&5&99.64\%\\
    \texttt{t469}&1551&252&252&99.50\%&236&236&8
    &99.52\%&236&236&6&99.52\%&214&40&5&99.50\%\\
    \hline
    avg. &713&118&118&99.71\%&109&109&8&99.73\%
    &109&109&5&99.73\%&93&17&4&99.71\%\\
    ratio & --  &+21.19\% & -- & -- &+14.68\% & -- & -- &  -- &+14.68\% & -- & -- & -- & 1.00 & --  & -- & -- \\
    \hline
\end{tabular}
}
\end{table*}

\subsection{Comparison with Previous TSV Fault Tolerance Planning Work}
\label{sec:etp}

%

We use simulated annealing-based multi-layer floorplanning \cite{chen2010multi} to generate the block floorplan and the f-TSV planning method in \cite{xu2017clustered} to generate f-TSV planning result as the input to the proposed fault-tolerance TSV planning framework. Based on the same f-TSV planning result, we run the flow in \cite{wang2015defect,xu2017clustered}, and the proposed heuristic based framework, respectively.
The experiment is tested on MCNC and GSRC benchmarks, including two MCNC circuits (\texttt{ami33} and \texttt{ami49}), and four GSRC circuits (\texttt{n50}, \texttt{n100}, \texttt{n200} and \texttt{n300}).
\revise{
We adopt one more industrial 2D design, which contains 403266 cells and 448514 nets.
{hMetis} \cite{karypis1999multilevel} is adopted to partition the design into several blocks for floorplanning.
Based on different block numbers, two benchmark cases, \texttt{t337} and \texttt{t469}, are generated.
That is, \texttt{t337} has 337 blocks and 1836 nets, while \texttt{t469} has 469 blocks and 5479 nets.
Since the square has the smallest perimeter among all the rectangles with the same area \cite{chen2008fixed}, here the shapes of all the blocks are set to square.
}
The experiment is executed 20 times independently for each benchmark.

In fault-tolerance structures, the multiplexers are used to reroute signals, and the delay of a multiplexer is increased along with the number of input ports.
Besides the hardware cost incurred by the fault-tolerance structure is related to the number of s-TSVs.
In this experiment, we compare the number of s-TSVs and the maximum port number of multiplexers of \cite{wang2015defect,xu2017clustered}, and the proposed TSV planning framework under 3-fault tolerance structures.
The layer number is set to 3. The target chip yield is set to 99.7\% and the TSV defect probability $p$ is set to 0.001.
The yield results in experiment are accurate to the fourth decimal place.
3 s-TSVs are assigned to each f-TSV group in \cite{wang2015defect,xu2017clustered}, that is, the maximum number of tolerant faults $K$ equals to 3.

TABLE~\ref{tab:tab7} lists the statistic results averaged over 20 independent experiments.
All results listed in table satisfy the target chip yield.
Column ``\#f-TSV'' represents the total number of f-TSVs. Since the three frameworks are run on the same f-TSV planning result, the number of f-TSVs is the same.
\revise{Columns} ``\#s-TSV'', ``\#gp'', and ``Yield'' list the total number of allocated s-TSVs, the number of groups, and the chip yield, respectively.
Besides, column ``\#Port'' provides the maximum port number of multiplexers among all groups, while column $K$ gives the number of tolerant faults in that group, respectively.
Since the generation of fault-tolerance structure is not considered in~\cite{wang2015defect}, the maximum port number of multiplexers is not listed.
As shown in TABLE~\ref{tab:tab7}, the number of f-TSV groups is greatly reduced in the proposed method.
Compared with \cite{wang2015defect} and \cite{xu2017clustered}, the proposed fault tolerance TSV planning framework can reduce the number of used s-TSVs by \textcolor{blue}{32.79\%} and \textcolor{blue}{31.67\%} on average, respectively.
In addition, in the proposed framework, if the maximum $K$ is used for each group, it will cause larger multiplexers.
Because the maximum number of tolerant faults ($K$) in adaptive fault-tolerance structures is often much greater than that of \cite{xu2017clustered}, which is fixed at $3$.
As a result, the maximum port number of multiplexers is increased accordingly in the generated fault-tolerance structures.

To reduce the size of required multiplexers, we also run the proposed fault tolerance TSV planning framework with $K\leq3$, that is, we set $K$ to $3$ if the maximum number of tolerant faults $K$ in a group is greater than $3$.
As shown in TABLE~\ref{tab:tab7}, compared with \cite{xu2017clustered}, the proposed fault tolerance TSV planning framework with $K\leq3$ has comparable maximum port number of multiplexers.
But the required s-TSVs are surprisingly reduced by \textcolor{blue}{50\%} on average under the same target yield, as shown in TABLE~\ref{tab:tab7}.

The TSV defect probability $p$ in~\cite{jiang2012effective} ranges from 0.001 to 0.01.
In order to see the impact of $p$ on performance, we also execute the experiment when $p$ is set to 0.01 under 3-fault tolerance structures.
The layer number is set to 3. The target chip yield is set to 99.5\%.
TABLE~\ref{tab:tab10} lists the statistic results averaged over 20 independent experiments.
All results listed in table satisfy the target chip yield.
Based on the same f-TSV planning result, we run the flow in \cite{wang2015defect,xu2017clustered}, and the proposed heuristic based framework, respectively.
Compared with \cite{wang2015defect} and \cite{xu2017clustered}, the proposed fault tolerance TSV planning framework can reduce the number of used s-TSVs by 32.24\% and 31.01\% on average, respectively.
In order to reduce the size of required multiplexers, we also run the proposed fault tolerance TSV planning framework with $K\leq3$.
As shown in TABLE~\ref{tab:tab10}, compared with \cite{xu2017clustered}, the proposed fault tolerance TSV planning framework with $K\leq3$ has comparable maximum port number of multiplexers.
But the required s-TSVs are surprisingly reduced by 46.50\% on average under the same target yield, as shown in TABLE~\ref{tab:tab10}.

Besides, in~\cite{chen2015novel}, $1$-fault tolerance structures are generated using minimum spanning tree based method. However, it is difficult to apply the method to the fault-tolerance structure using more than one spare TSVs.
In addition, the delay overhead introduced by the multiplexers, which are used for rerouting signals in the generated fault-tolerance structures, is not considered. In the worst-case the input port number of a multiplexer could be the number of f-TSVs in the group if the tree is a star structure,
which introduces large delay overhead.
In this experiment, we consider $1$-fault tolerance structures case, that is, the maximum number of tolerant faults $K$ equals to 1.
Since the chip yield is lower under $1$-fault tolerance structures, the target chip yield is set to 99.5\% and the TSV defect probability $p$ is set to 0.001.
And we compare \cite{chen2015novel}, \cite{wang2015defect}, \cite{xu2017clustered}, with the proposed heuristic based model under $1$-fault tolerance structures.
One s-TSV is assigned to each f-TSV group in \cite{wang2015defect} and \cite{xu2017clustered}. And we also set $K$ to $1$ in the proposed fault tolerance TSV planning framework, if the maximum number of tolerant faults $K$ in a group is greater than $1$.
Based on the TSV planning method in~\cite{xu2017clustered}, we run the minimum spanning tree method in~\cite{chen2015novel}.
Therefore, the s-TSV numbers and chip yield of~\cite{chen2015novel} and~\cite{xu2017clustered} are same in the experiment.

TABLE~\ref{tab:tab8} lists the statistic results averaged over 20 independent experiments.
As shown in TABLE~\ref{tab:tab8}, compared with \cite{chen2015novel} and \cite{xu2017clustered}, the proposed fault tolerance TSV planning framework can reduce the number of s-TSVs and the maximum port number of multiplexers when generating  $1$-fault tolerance structures.

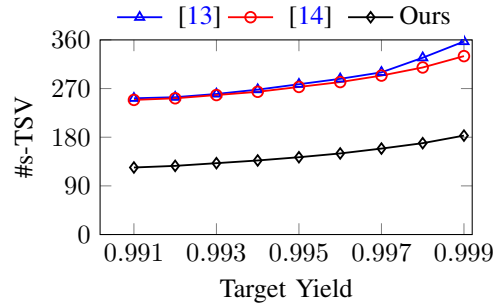
\begin{figure}[tb!]
  \centering
  \input{pgfplot/stsvvsty}
  \caption{The number of required s-TSVs under various target yields.}
  \label{fig:stsvvsty}
\end{figure}

Fig.~\ref{fig:stsvvsty} shows the \revise{required s-TSV numbers under various target yields, in comparison among \cite{wang2015defect}, \cite{xu2017clustered}, and our proposed framework.} 
The experiment is performed on \texttt{n100} benchmark.
Each data point in the figure is an average of 20 independent experiments.
It can be observed that the number of required s-TSVs increases \revise{along with increasing target yield and is significantly reduced by the proposed framework for all target chip yields}.


%% file: pgfplot/stsvvsty.tex
\definecolor{myorange}{RGB}{244,106,18} 
\definecolor{myblue}{RGB}{0,111,190}    
\definecolor{mygreen}{RGB}{0,127,128}   
\definecolor{myred}{RGB}{228,46,36}     
\definecolor{myyellow}{RGB}{198,148,34} 
\definecolor{mydark}{RGB}{114,44,114}   
\definecolor{mymiddle}{RGB}{144,44,144} 
\definecolor{mylight}{RGB}{167,44,167}  

\pgfplotsset{
    width =0.36\textwidth,
    height=0.23\textwidth
}
\begin{tikzpicture}
\begin{axis}[minor tick num=0,
xmax = 0.9992,
ymin=0, ymax=360,
xticklabel style={/pgf/number format/fixed,/pgf/number format/precision=3},
xtick={0.991,0.993,...,1.000},
ytick={0,90,...,360},
yticklabel style={/pgf/number format/fixed,/pgf/number format/precision=0},
xlabel={Target Yield},
ylabel={\#s-TSV},
xlabel near ticks,
legend style={
  draw=none,
  at={(0.50,1.23)},
  anchor=north,
  legend columns=3,
}
]
\addplot +[line width=0.7pt] [color=blue,  solid, mark=triangle]  table [x={targetyield},  y={iccad}]  {stsvvsty.dat};
\addplot +[line width=0.7pt] [color=red,   solid, mark=o]         table [x={targetyield},  y={tcad}]  {stsvvsty.dat};
\addplot +[line width=0.7pt] [color=black, solid, mark=diamond]   table [x={targetyield},  y={proposed}]  {stsvvsty.dat};
\legend{{\cite{wang2015defect}},{\cite{xu2017clustered}}\ \ ,Ours }
\end{axis}
\end{tikzpicture}

%% file: doc/conclu.tex
\section{Conclusion}
\label{sec:conclu}

In this paper, we focus on the generation of adaptive TSV fault-tolerance structure.
An integer linear programming (ILP) based model and an efficient min-cost-max-flow based heuristic method are proposed to generate the adaptive fault-tolerance structures in minimizing both the multiplexer delay overhead and the used s-TSV number.
In the end, a fault-tolerance TSV planning methodology is also proposed to provide yield awareness in TSV planning.
Experimental results show that, compared with state-of-the-art, the proposed fault tolerance TSV planning methodology can effectively reduce the number of s-TSVs used for fault tolerance.

{\color{blue}
Besides, in this work, the proposed TSV fault tolerance planning is performed in floorplanning stage and we have no accurate timing information.
Therefore, we only use the wirelength to reflect the wire delay in floorplanning stage.
In future we plan to evaluate the delay more accurately by executing time-consuming routing.
}

\section*{Acknowledgments}
The authors would like to thank the Information Science Laboratory Center of USTC for hardware and software services.

%% file: TCAD-TSV-Flow.bbl

%% file: TCAD-TSV-Flow.bbl
\begin{thebibliography}{10}
\providecommand{\url}[1]{#1}
\csname url@samestyle\endcsname
\providecommand{\newblock}{\relax}
\providecommand{\bibinfo}[2]{#2}
\providecommand{\BIBentrySTDinterwordspacing}{\spaceskip=0pt\relax}
\providecommand{\BIBentryALTinterwordstretchfactor}{4}
\providecommand{\BIBentryALTinterwordspacing}{\spaceskip=\fontdimen2\font plus
\BIBentryALTinterwordstretchfactor\fontdimen3\font minus
  \fontdimen4\font\relax}
\providecommand{\BIBforeignlanguage}[2]{{%
\expandafter\ifx\csname l@#1\endcsname\relax
\typeout{** WARNING: IEEEtran.bst: No hyphenation pattern has been}%
\typeout{** loaded for the language `#1'. Using the pattern for}%
\typeout{** the default language instead.}%
\else
\language=\csname l@#1\endcsname
\fi
#2}}
\providecommand{\BIBdecl}{\relax}
\BIBdecl

\bibitem{souri2000multiple}
S.~J. Souri, K.~Banerjee, A.~Mehrotra, and K.~C. Saraswat, ``{Multiple Si layer
  {ICs}: Motivation, performance analysis, and design implications},'' in
  \emph{ACM/IEEE Design Automation Conference (DAC)}, 2000, pp. 213--220.

\bibitem{joyner2001global}
J.~W. Joyner, P.~Zarkesh-Ha, and J.~D. Meindl, ``A global interconnect design
  window for a three-dimensional system-on-a-chip,'' in \emph{IEEE
  International Interconnect Technology Conference (IITC)}, Jun. 2001, pp.
  154--156.

\bibitem{ITRS}
``International technology roadmap for semiconductors,''
  [Online].\url{http://www.itrs2.net}.

\bibitem{tcad2017Lu}
T.~Lu, C.~Serafy, Z.~Yang, S.~K. Samal, S.~K. Lim, and A.~Srivastava,
  ``\textcolor{blue}{{TSV}-Based 3-D ICs: Design Methods and Tools},''
  \emph{IEEE Transactions on Computer-Aided Design of Integrated Circuits and
  Systems (TCAD)}, vol.~36, no.~10, pp. 1593--1619, 2017.

\bibitem{loi2008low}
I.~Loi, S.~Mitra, T.~H. Lee, S.~Fujita, and L.~Benini, ``A low-overhead fault
  tolerance scheme for {TSV}-based {3D} network on chip links,'' in
  \emph{IEEE/ACM International Conference on Computer-Aided Design (ICCAD)},
  Nov. 2008, pp. 598--602.

\bibitem{chen2015novel}
Y.-G. Chen, W.-Y. Wen, Y.~Shi, W.-K. Hon, and S.-C. Chang, ``Novel spare {TSV}
  deployment for {3-D ICs} considering yield and timing constraints,''
  \emph{IEEE Transactions on Computer-Aided Design of Integrated Circuits and
  Systems (TCAD)}, vol.~34, no.~4, pp. 577--588, 2015.

\bibitem{xu2012yield}
Q.~Xu, L.~Jiang, H.~Li, and B.~Eklow, ``Yield enhancement for {3D}-stacked
  {ICs}: Recent advances and challenges,'' in \emph{IEEE/ACM Asia and South
  Pacific Design Automation Conference (ASPDAC)}, Feb. 2012, pp. 731--737.

\bibitem{lee2009test}
H.-H.~S. Lee and K.~Chakrabarty, ``Test challenges for {3D} integrated
  circuits,'' \emph{IEEE Design \& Test of Computers}, vol.~26, no.~5, pp.
  26--35, 2009.

\bibitem{ferri2007strategies}
C.~Ferri, S.~Reda, and R.~I. Bahar, ``Strategies for improving the parametric
  yield and profits of {3D ICs},'' in \emph{IEEE/ACM International Conference
  on Computer-Aided Design (ICCAD)}, Nov. 2007, pp. 220--226.

\bibitem{chou2010yield}
C.-W. Chou, Y.-J. Huang, and J.-F. Li, ``Yield-enhancement techniques for {3D}
  random access memories,'' in \emph{International Symposium on VLSI Design,
  Automation, and Test (VLSI-DAT)}, Apr. 2010, pp. 104--107.

\bibitem{jiang2010yield}
L.~Jiang, R.~Ye, and Q.~Xu, ``Yield enhancement for {3D}-stacked memory by
  redundancy sharing across dies,'' in \emph{IEEE/ACM International Conference
  on Computer-Aided Design (ICCAD)}, Nov. 2010, pp. 230--234.

\bibitem{jiang2012effective}
L.~Jiang, Q.~Xu, and B.~Eklow, ``On effective {TSV} repair for {3D}-stacked
  {ICs},'' in \emph{IEEE/ACM Proceedings Design, Automation and Test in Eurpoe
  (DATE)}, Mar. 2012, pp. 793--798.

\bibitem{wang2015defect}
S.~Wang, M.~B. Tahoori, and K.~Chakrabarty, ``Defect clustering-aware
  spare-{TSV} allocation for {3D ICs},'' in \emph{IEEE/ACM International
  Conference on Computer-Aided Design (ICCAD)}, Nov. 2015, pp. 307--314.

\bibitem{xu2017clustered}
Q.~Xu, S.~Chen, X.~Xu, and B.~Yu, ``Clustered fault tolerance {TSV} planning
  for {3D} integrated circuits,'' \emph{IEEE Transactions on Computer-Aided
  Design of Integrated Circuits and Systems (TCAD)}, vol.~36, no.~8, pp.
  1287--1300, 2017.

\bibitem{chen2010cost}
Y.~Chen, D.~Niu, Y.~Xie, and K.~Chakrabarty, ``Cost-effective integration of
  three-dimensional {(3D) ICs} emphasizing testing cost analysis,'' in
  \emph{IEEE/ACM International Conference on Computer-Aided Design (ICCAD)},
  Nov. 2010, pp. 471--476.

\bibitem{brandon2014design}
B.~Noia and K.~Chakrabarty, \emph{Design-for-Test and Test Optimization
  Techniques for {TSV}-based {3D} Stacked {ICs}}.\hskip 1em plus 0.5em minus
  0.4em\relax Switzerland: Springer, 2014.

\bibitem{jiang2013effective}
L.~Jiang, Q.~Xu, and B.~Eklow, ``On effective through-silicon via repair for
  {3-D} stacked {ICs},'' \emph{IEEE Transactions on Computer-Aided Design of
  Integrated Circuits and Systems (TCAD)}, vol.~32, no.~4, pp. 559--571, 2013.

\bibitem{schrijver}
A.~Schrijver, \emph{{Combinatorial Optimization: Polyhedra and
  Efficiency}}.\hskip 1em plus 0.5em minus 0.4em\relax Berlin: Springer Science
  \& Business Media, 2002, vol.~24.

\bibitem{TOOL_B1999_leda}
K.~Mehlhorn and S.~Naher, \emph{{LEDA: A Platform for Combinatorial and
  Geometric Computing}}.\hskip 1em plus 0.5em minus 0.4em\relax Cambridge
  University Press, 1999.

\bibitem{TOOL_glpk}
A.~Makhorin, ``{GLPK (GNU linear programming kit)},'' 2008.

\bibitem{karypis1999multilevel}
G.~Karypis, R.~Aggarwal, V.~Kumar, and S.~Shekhar, ``Multilevel hypergraph
  partitioning: applications in {VLSI} domain,'' \emph{IEEE Transactions on
  Very Large Scale Integration Systems (TVLSI)}, vol.~7, no.~1, pp. 69--79,
  1999.

\bibitem{chen2010multi}
S.~Chen and T.~Yoshimura, ``Multi-layer floorplanning for stacked {ICs}:
  Configuration number and fixed-outline constraints,'' \emph{Integration, the
  VLSI Journal}, vol.~43, no.~4, pp. 378--388, 2010.

\bibitem{chen2008fixed}
------, ``Fixed-outline floorplanning: Block-position enumeration and a new
  method for calculating area costs,'' \emph{IEEE Transactions on
  Computer-Aided Design of Integrated Circuits and Systems (TCAD)}, vol.~27,
  no.~5, pp. 858--871, 2008.

\end{thebibliography}
